\documentclass[aps,twocolumn,prd,showpacs,nofootinbib]{revtex4}
\usepackage{amsmath}
\usepackage{graphicx}
\usepackage{dcolumn}
\usepackage{bm}
\usepackage{amssymb}
\usepackage{latexsym}
\usepackage{color}

\def\be{\begin{equation}}
\def\ee{\end{equation}}
\def\bea{\begin{eqnarray}}
\def\eea{\end{eqnarray}}

\begin{document}

\title{Curvaton with Nonminimal Derivative Coupling to Gravity II: Full Perturbation Analysis}

\author{Kaixi Feng$^1$}
\email{fengkaixi10@mails.gucas.ac.cn}

\author{Taotao Qiu$^{2,3}$}
\email{qiutt@mail.ccnu.edu.cn}


\vspace{16mm}

\affiliation{$1$ School of Physics, University of Chinese Academy of Sciences, Beijing 100049, China \vspace{1mm}}
\affiliation{$2$ Institute of Astrophysics, Central China Normal University, Wuhan 430079, China}
\affiliation{$3$ State Key Laboratory of Theoretical Physics, Institute of Theoretical Physics, Chinese Academy of Sciences, Beijing 100190, China}

\begin{abstract}
In our previous work \cite{Feng:2013pba}, we have shown a curvaton model where the curvaton has a nonminimal derivative coupling to gravity. Such a coupling could bring us scale-invariance of the perturbations for wide range constant values of the equation-of-state of the cosmic background at the early time. In this paper, we continue our study by fully analyzing its perturbations up to the third order. Apart from the usual 2-point correlation function that has already been calculated in \cite{Feng:2013pba}, we have also taken into account the 3-point correlation functions including pure scalar part, pure tensor part, as well as the cross-correlations between scalar and tensor perturbation modes. We find that for pure scalar part, the 3-point correlation functions can generate non-Gaussianities that fits the PLANCK data very well. For pure tensor and mixed parts, the shape functions have peaks at squeezed and equilateral limits respectively, responsible for sizable $f_{NL}^{sqz}$ and $f_{NL}^{eql}$, which could be tested by the future observatioanl data.

\end{abstract}

\maketitle

\section{introduction}

Curvaton has been widely discussed in the literature as an alternative of inflaton to generate primordial perturbations in the early universe \cite{Lyth:2001nq,Mollerach,Linde:1996gt, Enqvist:2001zp, Moroi:2001ct}. During inflation period, with the addition of the so-called ``curvaton field" which can generate the most of the primordial perturbations needed in the early times, constraints on the background evolution of the universe can be widely released, and it is easier to have viable models which can fit the data. At the end of inflation, the (isocurvature) perturbations generated by curvaton can be transferred into adiabatic ones \cite{Lyth:2001nq}, which is required by the observations. For extended study of curvaton models, see \cite{Sasaki:2006kq,Allahverdi:2006dr,DBIcurvaton,runningfnl,Cai:2011zx,Qiu:2011cy,higgs,Li:2013hga}.

Recently, we studied a new curvaton model with its kinetic term coupled nonminimally to the Einstein tensor, $G_{\mu\nu}$. This kind of coupling can be viewed as a subgroup of Horndeski theory \cite{Horndeski:1974wa}, or the most general scalar-tensor theory \cite{Deffayet:2009mn}, which is regarded as a ghost-free theory even when null energy condition is violated. Moreover, due to such kind of coupling, we show that the scale-dependence of the scalar perturbations can be independent of the background evolution, and is scale-invariant, provided only that the background equation-of-state is nearly a constant. That means, the scale-invariant power spectrum can be obtained in this model, even if the background is not inflation at all. We also take into account the tensor perturbation. although it can bring some constraints on the value of EoS, still a large range of value is allowed. Furthermore, we also discussed the transfer from curvaton perturbations into curvature perturbations , as well as local non-Gaussianity generated in this model.

In early 2013, the PLANCK satellite released its first observational result about cosmology, and one highlight point is the accurate but small non-Gaussianity of the early universe \cite{Ade:2013zuv}. The PLANCK data shows that the estimator of the equilateral non-Gaussianity is within $-42\pm75$ (1$\sigma$), while that of squeezed ones is within $2.7\pm5.8$ (1$\sigma$). It is quite an attracting result, which could rule out many early universe models. Although we have shown that the local non-Gaussianity generated in our model is consistent with the data, full non-Gaussianity analysis including the equilateral ones are still not taken. As a completion of the study of this model, in this paper we will study the other shapes of its non-Gaussianities to see if this model can pass the newest observational data.

This paper is organized as follows: in Sec. II we briefly review the background evolution of our model, and in Sec. III we analyze its perturbations. After given the basic perturbation equations, in the second and third subsections we show the results of perturbation at the second order. From subsection {\bf D} to subsection {\bf G}, we calculated the third order perturbations of pure tensor part, pure scalar part, one tensor plus two scalar part and one scalar plus two tenors part, respectively. In each part, we plot the shape function of the correlation functions, and especially for pure scalar part, we give the formulation of the non-Gaussianity estimator $f_{NL}$ in its equilateral limit in order to compare with the observational constraints by PLANCK data. The last section is our conclusion.

\section{The model: background equations}
The action of nonminimal derivative coupling curvaton is considered as \cite{Feng:2013pba}
\be\label{action}
\mathcal{S}=\int\mathrm{d}^4x\sqrt{g}\Big[\frac{R}{16\pi G}+\frac{\xi}{M^2}G_{\mu\nu}\partial^\mu\varphi\partial^\nu\varphi-V(\varphi)+{\cal L}_{bg}\Big]~,
\ee
where $G_{\mu\nu}$ is the Einstein tensor: $G_{\mu\nu}\equiv R_{\mu\nu}-g_{\mu\nu}R/2$, and $\xi$ is an arbitrary coefficient. The nonminimal derivative coupling field was first proposed in \cite{Amendola:1993uh} and has been analyzed in various aspects of cosmology and gravity theories \cite{Capozziello:1999xt,Granda:2011zk,Germani:2010gm,Sadjadi:2012zp,Capozziello:1999uwa,Granda:2009fh,Saridakis:2010mf,Granda:2011eh,Sadjadi:2010bz, Banijamali:2011qb,Banijamali:2012kq,Chen:2010ru,Lin:2011zzd,Rinaldi:2012vy,Cartier:2001is,Sushkov:2009hk,Daniel:2007kk,Gao:2010vr,Anabalon:2013oea}. In \cite{Deffayet:2009mn}, Deffayet et, al. showed that field with nonminimal derivative coupling can be included into the generalized Galileon theory which is inspired from the Horndeski theory \cite{Horndeski:1974wa}, and it can have the appealing property that there will be no ghost modes. Another curvaton model made of Galileon is given in \cite{Wang:2011dt}, where the full scalar perturbations has been calculated.

It is straightforward to write down the equation of motion for the curvaton field $\varphi$, such as
\be\label{eom}
\frac{6\xi}{M^{2}}H^{2}\ddot{\varphi}+\frac{6\xi}{M^{2}}(2\dot{H}+3H^{2})H\dot{\varphi}+V_{\varphi}=0~,
\ee
and its energy density and pressure can be expressed as
\bea\label{rho}
\rho_\varphi&=&\frac{9\xi}{M^2}H^2\dot\varphi^2+V(\varphi)~,\\
\label{p} P_\varphi&=&-\frac{\xi}{M^2}(3H^2\dot\varphi^2+2\dot H\dot\varphi^2+4H\dot\varphi\ddot\varphi)-V(\varphi)~,
\eea
respectively. The background evolution of the curvaton field with various types of potential has been classified and briefly analyzed in \cite{Feng:2013pba}. Moreover, we can define the following parameters:
\be\label{para} y\equiv\frac{\xi}{M^{2}}\dot{\varphi}^{2}~,~\eta\equiv\frac{\ddot{\varphi}}{H\dot\varphi}~,~\epsilon\equiv-\frac{\dot{H}}{H^{2}}~,~\epsilon_{\phi}\equiv\frac{\dot{\phi}^{2}}{M_{p}^{2}H^{2}}~,
\ee
for later convenience. For inflationary background, $|\eta|$, $|\epsilon|$, $|\epsilon_\phi|$ are much smaller than unity. Notice also that $\dot{y}=2\xi\dot{\varphi}\ddot{\varphi}/M^{2}=2Hy\eta$.

\section{The model: perturbation analysis}
\subsection{basic equations}
In this subsection, we give the basic formulations of perturbation of our model up to 3rd order which will be used in the following subsections. First of all,the perturbed metric can be written as:
\be\label{adm}
ds^{2}=-N^{2}dt^{2}+h_{ij}(dx^{i}+N^{i}dt)(dx^{j}+N^{j}dt)~,
\ee
where $N$ is the lapse function, $N^i$ is the shift vector, and $h_{ij}$ is the induced 3-metric. One can then perturb these functions as:
\be\label{metricpert}
N=1+\alpha~,~N_i=\partial_i\beta~,~h_{ij}=a^2(t)e^{2\psi}e^{\gamma_{ij}}~,
\ee
where $\alpha$, $\beta$ and $\psi$ are the scalar metric perturbations, and $e^{\gamma_{ij}}\equiv\delta_{ij}+\gamma_{ij}+(1/2)\gamma_i^k\gamma_{kj}+(1/6)\gamma_i^k\gamma_k^l\gamma_{lj}+...$ denotes the tensor part of perturbation. The pertubation of $\varphi$ field is
\be\label{fieldpert}
\varphi\rightarrow \varphi(t)+\delta\varphi(t,\bf{x})~.
\ee
In the following, we will take the spatial-flat gauge $\psi=0$ for convenience, and neglect the perturbation of the background field $\phi$.

Using these perturbation elements, we could then expand the curvaton action (\ref{action}) order by order:
\be\label{perturbaction} \mathcal{S}=\mathcal{S}_0+\mathcal{S}_{ss}+\mathcal{S}_{tt}+\mathcal{S}_{sss}+\mathcal{S}_{ttt}+\mathcal{S}_{stt}+\mathcal{S}_{sst}+...~,
\ee
where the subscript ``0" denotes background, $``ss"$ and $``tt"$ means scalar and tensor parts for 2nd order perturbation, and $``sss"$, $``ttt"$, $``stt"$ and $``sst"$ are the pure scalar part, pure tensor part, two scalar coupled with one tensor part, and one scalar coupled with two scalar part of 3rd order perturbations, respectively. As is well known that at the 2nd order, the scalar and tensor modes decouples with each other, while from the 3rd order, they are coupled together and we should also take into account their cross correlation function. Perturbations equal to (or higher than) 4th order are regarded as negligible corrections and will not be considered in this context.

The n-point correlation functions of any perturbative quantity (in its momentum space) $\delta({\bf k},t)$ is defined as:
\be
\langle\delta({\bf k}_1,t)\delta({\bf k}_2,t)...\delta({\bf k}_n,t)\rangle
\ee
where $\delta({\bf k},t)$ can either be the scalar purturbation $\delta\varphi_k({\bf k},t)$ or tensor perturbation $\gamma^{(k)}_{ij}({\bf k},t)$. Specifically, the two-point correlation functions is
\be
\langle\delta({\bf k}_1,t)\delta({\bf k}_2,t)\rangle=|\delta(k_1,t)|^2\delta^3({\bf k}_1+{\bf k}_2)~
\ee
where $k_1=|{\bf k}_1|$, and the three-point correlation functions is
\bea\label{ng}
&&\langle\delta({\bf k}_1,t)\delta({\bf k}_2,t)\delta({\bf k}_3,t)\rangle\nonumber\\
&=&-i\langle|\int_{t_i}^{t_f}dt^\prime [ \delta({\bf k}_1,t)\delta({\bf k}_2,t)\delta({\bf k}_3,t),H^{(3)}_{int}(\delta,t^\prime)]|\rangle~,
\eea
where $H^{(3)}_{int}$ is the Hamiltonian extracted from the 3rd order perturbed action of (\ref{perturbaction}).

We observe these correlation functions by their spectra. The spectrum for two point correlation function (power spectrum) is defined as:
\be
{\cal P}_{\delta\ast}(k)\equiv\frac{k^3}{2\pi^2}|\delta(k,t=t_\ast)|^2=A_s\left(\frac{k}{k_*}\right)^{n_s-1}~,
\ee
For power spectrum, it is important to know its amplitude $A_s$, as well as its spectral index $n_s$ which describes dependence with the wavenumber of the perturbation modes, $k$, since they can directly be connected with observational data. Recent PLANCK 2013 data gives the constraint of $A_s=(2.23\pm0.16)\times 10^{-9}$, $n_s=0.9603\pm0.0073$ ($1\sigma$). While the spectrum for three correlation function (bispectrum) is defined as:
\be\label{correlation}
{\cal B}({\bf k}_1,{\bf k}_2,{\bf k}_3)=\frac{\langle\delta({\bf k}_1,t)\delta({\bf k}_2,t)\delta({\bf k}_3,t)\rangle}{(2\pi)^3\delta^3({\bf k}_1+{\bf k}_2+{\bf k}_3)}\bigg|_{t=t_\ast}~.
\ee

For bispectrum, we take care of the shape of the correlation function as functions of wavenumbers of each point, ${\cal A}({\bf k}_1,{\bf k}_2,{\bf k}_3)$, which is defined via bispectrum as:
\be\label{shape}
{\cal B}({\bf k}_1,{\bf k}_2,{\bf k}_3)=\frac{(2\pi)^4{\cal P}_{\delta}^2}{\prod_{i=1}^3k_i^3}{\cal A}({\bf k}_1,{\bf k}_2,{\bf k}_3)~.
\ee
Moreover, one can also define the estimator,
\be\label{fnl}
f_{NL}=\frac{10}{3}\frac{{\cal A}({\bf k}_1,{\bf k}_2,{\bf k}_3)}{\sum_{i=1}^3k_i^3}~.
\ee
which can be constrained directly by the observational data. The PLANCK data have imposed stringent constraints on both equilateral and squeezed limits of $f_{NL}$, namely $f_{NL}^{eql}=-42\pm75$ and $f_{NL}^{sqz}=2.7\pm5.8$ (1$\sigma$). These results can be well used to constraint models that give rise to primordial non-Gaussianities.

The spectrum for four point correlation function is called trispectrum, and so on and so forth. The trispectrum is described by its shape as well as the estimator $g_{nl}$ and $\tau_{nl}$. Up till now, the constraints on trispectum is still very poor, only having an upper bound of $\tau_{nl}<2800$ (2$\sigma$).
\subsection{Two-point correlation function: scalar part}
In this subsection, we focus on the scalar perturbation up to 2nd order in this model, which is basically obtained from $\mathcal{S}_{ss}$ in (\ref{perturbaction}). First of all, we notice that $N$ and $N_i$ in (\ref{adm}) are only constraint quantities and have no dynamics, we can make use of techniques in \cite{Chen:2006nt} to express them using field variables:
\be
\alpha=a_1\dot{\delta\varphi}+a_2\delta\varphi~,~~~\partial^2\beta=b_1\dot{\delta\varphi}+b_2\delta\varphi+b_3\partial^2\delta\varphi~,
\ee
where we define
\bea
a_1&\equiv&-\frac{2\xi\dot{\varphi}/M^2}{M_p^2-3y}~,~a_2\equiv\frac{3\xi H\dot{\varphi}/M^2}{M_p^2-3y}~,\\
b_1&\equiv&a^{2}\frac{[a_{1}\frac{\dot{\phi}^{2}}{2H}-\frac{9\xi H\dot{\varphi}}{M^{2}}-3H a_{1}(M_{p}^{2}-6y)]}{M_p^2-3y}~,\\
b_2&\equiv&a^{2}\frac{[a_{2}\frac{\dot{\phi}^{2}}{2H}-3H a_{2}(M_{p}^{2}-6y)-\frac{V_{,\varphi}}{2H}]}{M_p^2-3y}~,\\
b_3&\equiv&-\frac{2\xi\dot{\varphi}/M^{2}}{M_{p}^{2}-3y}~.
\eea

Then the next step is to expand the action to second order of $\delta\varphi$. As has been demonstrated in \cite{Feng:2013pba}, we choose the kinetic term of the field $\varphi$ to be $\sim G_{\mu\nu}\partial^\mu\varphi\partial^\nu\varphi$ such that the 2nd order perturbation action will be like
\be\label{perturb}
\mathcal{S}_{ss}=\int d^3xd\eta\frac{a^2Q_s}{c_s^2}\left[{\delta\varphi^\prime}^2-c_s^2(\partial \delta\varphi)^2\right]~,
\ee
where
\be
Q_s\equiv P_{,X}\sim\frac{1}{a^2(\eta_*-\eta)^2}~,~~~c_s^2\equiv\frac{P_{,X}}{\rho_{,X}}~,
\ee
and $^\prime$ means derivative with respect to conformal time $\eta$. However, since we worry that the nonminimal coupling will bring nontrivial effects to the perturbation action, we fully expand the action without neglecting the metric perturbation. We then obtain a more complete form of perturbed action:
\be\label{perturb2}
\mathcal{S}_{ss}=\int d\eta  d^{3}xa^2\frac{Q_s}{c_s^2}\Big[{\delta\varphi}^{\prime 2}-c_s^2\partial_{i}\delta\varphi\partial^{i}\delta\varphi-\frac{1}{2}\frac{a^2c_s^2m_{eff}^{2}}{Q_s}\delta\varphi^{2}\Big]~,
\ee
where
\be\label{smally}
Q_s\simeq-\frac{\xi H^{2}}{M^{2}}(2\epsilon-7)~,~c_{s}^{2}\simeq-\frac{1}{3}(2\epsilon-7)~,~m_{eff}^{2}\simeq V_{\varphi\varphi}~.
\ee
in the $|y|\ll M_{p}^{2}$ limit while their full expression is given in Eqs. (20) of \cite{Feng:2013pba}. In order to make this model free of ghost and gradient instabilities, we require $Q_s>0$, $c_s^2>0$, which leads to
\be\label{constraint1}
\xi>0~,~~~w=-1+\frac{2}{3}\epsilon<\frac{4}{3}~,
\ee
which is the region of viability of our model in this case. From the action (\ref{perturb2}), one can get the equation of motion for $\delta\varphi$. Define $z_s\equiv a\sqrt{Q_s}/c_s$, one have:
\be\label{eompertscalar}
(z_s\delta\varphi)^{\prime\prime}+(c_s^2k^2-\frac{z_s^{\prime\prime}}{z_s}+\frac{1}{2}\frac{a^4m_{eff}^2}{z_s^2})(z_s\delta\varphi)=0~.
\ee

It is convenient to solve the equation in momentum space. The variables $\delta\varphi$ in its momentum space are:
\bea\label{fourierscalar}
\delta\varphi({\bf x},\eta)&=&\int d^3k \delta\varphi({\bf k},\eta)e^{i{\bf k}\cdot{\bf x}}~,\\
\delta\varphi({\bf k},\eta)&=&\delta\varphi_k(\eta)\alpha_s({\bf k})+\delta\varphi_k^\ast(\eta)\alpha_s^\dagger(-{\bf k})~,
\eea
where $\alpha_s({\bf k})$ and $\alpha_s^\dagger({\bf k})$ are producing and annihilating operators satisfying the commutation relation $[\alpha_s({\bf k}),\alpha_s^\dagger({\bf k}^\prime)]=(2\pi)^3\delta({\bf k}-{\bf k}^\prime)$.

Substituting (\ref{fourierscalar}) into Eq. (\ref{eompertscalar}) and impose the initial condition of Bunch-Davies vacuum, one have
\be\label{scalarsub}
\delta\varphi_k=\frac{iH}{\sqrt{2Q_sc_sk^3}}(1+ic_{s}k\eta)e^{-ic_{s}k\eta}
\ee
for sub-horizon region and
\be\label{scalarsuper}
\delta\varphi_k\sim k^\nu(\eta_*-\eta)^{\nu+\frac{3}{2}}~,~~~k^{-\nu}(\eta_*-\eta)^{\nu+\frac{3}{2}}~,~\nu\simeq \frac{3}{2}~,
\ee
for super-horizon region, where we neglected the mass term. Finally, the (normalized) power spectrum of $\delta\varphi$ can be obtained as
\bea\label{spectrum}
\tilde{\cal P}_{\delta\varphi}&\equiv&\frac{{\cal P}_{\delta\varphi}}{M_p^2}=\frac{k^3}{2\pi^2}\frac{|\delta\varphi|^2}{M_p^2}~\nonumber\\
&=&\frac{H^2}{4\pi^2M_p^2c_s Q_s}=\sqrt{\frac{3}{(7-2\epsilon)^3}}\frac{M^{2}}{4\pi^2M_p^2\xi}~.
\eea
If the curvaton $\varphi$ has no potential, we will get an exact scale-invariant power spectrum, however, a slight tilt will arise if we allow a small potential which satisfies $V_{\varphi\varphi}\sim H^4\sim t^{-4}$.

The curvature perturbation can be generated either after the curvaton dominates or when the curvaton reaches equilibrium with the background. In Ref. \cite{Feng:2013pba}, we discussed the spectrum of the curvature perturbations in our model, and obtained results for both of the two mechanisms, which are:
\be
{\cal P}^{A}_\zeta\simeq\frac{27\sqrt{3}H_*^{2}}{4\pi^{2}|y_*|(7-2\epsilon)^{5/2}}\Big(\frac{\epsilon}{3+\epsilon-2\eta}\Big)^{2}~
\ee
and
\be
{\cal
P}^{B}_\zeta\simeq\frac{3\sqrt{3}\epsilon^{2}r^{2}H_*^{2}}{16\pi^{2}|y_*|(7-2\epsilon)^{5/2}}~
\ee
at the $|y|\ll M_p^2$ limit.

\subsection{Two-point correlation function: tensor part}
Besides scalar type of perturbations, the primordial perturbations of tensor type may also be generated in the early universe, which is expected to be detected by the coming PLANCK 2yr data. In this section, we will focus on the tensor part of the perturbations in our model. Consider only the tensor part of the perturbed metric in (\ref{adm}), one could easily obtain the tensor part of the 2nd order perturbed action as:
\be\label{perturbtensor}
\mathcal{S}_{tt} = \frac{1}{8}\int d\eta d^3x a^2\frac{Q_T}{c_T^2}[\gamma_{ij}^{\prime2}-c_T^2(\partial\gamma_{ij})^2]~,
\ee
where we defined
\be\label{QTcT}
Q_T=M_p^2+y~,~~~c_T^2=\frac{M_p^2+y}{M_p^2-y}~,
\ee
and in $|y|\ll M_p^2$ limit they will reduce to $Q_T\simeq M_p^2$, $c_T^2\simeq 1$.

One can solve the equation of motion for $\gamma_{ij}$ from the action (\ref{perturbtensor}) to get the tensor spectrum. Define $z_T\equiv a\sqrt{Q_T}/c_T$ and according to the action (\ref{perturbtensor}), one can have the equation of motion for $\gamma_{ij}$ as:
\be\label{eomperttensor}
(z_T\gamma_{ij})^{\prime\prime}+(c_T^2k^2-\frac{a^{\prime\prime}}{a})(z_T\gamma_{ij})=0~.
\ee

Similar as the scalar perturbation, the tensor perturbation can also be transformed into momentum space via Fourier transformation, which is:
\bea\label{fouriertensor}
\gamma_{ij}({\bf x},\eta)&=&\int d^3k \gamma_{ij}({\bf k},\eta)e^{i{\bf k}\cdot{\bf x}}~,\\
\gamma_{ij}({\bf k},\eta)&=&\sum_{\lambda=1}^2\Big[e_{ij}({\bf k},\lambda)\alpha_T({\bf k},\lambda)\gamma_k(\eta)\nonumber\\
       &&+e_{ij}^{\ast}(-{\bf k},\lambda)\alpha_T^\dagger(-{\bf k},\lambda)\gamma_k^\ast(\eta)\Big]~,
\eea
where $\alpha_T({\bf k},\lambda)$ and $\alpha_T^\dagger({\bf k},\lambda)$ are producing and annihilating operators, satisfying the commutation relation $[\alpha_T({\bf k},\lambda),\alpha_T^\dagger({\bf k}^\prime,\lambda^\prime)]=(2\pi)^3\delta_{\lambda\lambda^\prime}\delta({\bf k}-{\bf k}^\prime)$ and $e_{ij}$ is the polarization tensor with relations:
\bea\label{poltensor}
&e_{ii}({\bf k},\lambda)=0~,k^ie_{ij}({\bf k},\lambda)=0~,e_{ij}({\bf k},\lambda)e_{ij}({\bf k},\lambda^\prime)=\delta_{\lambda\lambda^\prime}~,&\nonumber\\
&e_{ij}^\ast({\bf k},\lambda)=e_{ij}(-{\bf k},\lambda)=e_{ij}({\bf k},-\lambda)~,&
\eea
and $\lambda=\pm2$. Solving Eq. (\ref{eomperttensor}), one has:
\be\label{tensorsub}
\gamma_k(\eta)=\frac{iH}{\sqrt{2Q_Tc_Tk^3}}(1+ic_Tk\eta)e^{-ic_Tk\eta}
\ee
for sub-horizon region and
\be\label{tensorsuper}
\gamma_{ij}\sim\text{constant.},~~~~\int\frac{dt}{a^3(t)M_p^2}~,
\ee
for super-horizon region. The power spectrum for tensor perturbation thus can be obtained as:
\be\label{spectrumtensor}
{\cal P}_T\sim k^{3}|\gamma_{ij}|^2\sim
\frac{H^2}{M_p^2}\big(\frac{k}{k_0}\big)^{n_T}~,
\ee
where $k_0$ denotes some pivot wavenumber. The spectral index
\bea
n_T&=&\frac{6(1+w)}{1+3w}~~~~(w>1)~,\nonumber\\
&or&\frac{12w}{1+3w}~~~~(-\frac{1}{3}<w<1)~
\eea
for contracting phase and
\be
n_T=\frac{6(1+w)}{1+3w}~~~~(w<-\frac{1}{3})~
\ee
for expanding phase. One can expect future PLANCK or BICEP data to put further constraints on the $n_T$.

\subsection{Three-point correlation function: pure scalar part}
In the following, we will discuss about 3-point correlation functions of our model, namely non-Gaussianities. Since for more than second order, the tensor perturbations couples to the scalar ones, the correlation functions contain not only pure scalar and tensor parts, but also have mixed parts between scalar and tensor modes. As a full investigation, we will analyze all these cases in the following subsection. First of all, we focus on the non-Gaussianities of pure scalar part, $\langle\delta\varphi\delta\varphi\delta\varphi\rangle$. Following (\ref{adm},\ref{metricpert},\ref{fieldpert}), one can get the 3rd-order perturbative action for the scalar part as:
\begin{widetext}
\bea\label{actionsss}
S_{sss}&\subset&\int dt L^{(3)}_{sss}~\nonumber\\
L^{(3)}_{sss}&=&a^3\Big[{\cal A}_1\dot{\delta\varphi}^3+{\cal A}_2a^{-2}\delta\varphi^2\partial^2\delta\varphi
    +{\cal A}_3a^{-2}\delta\varphi\partial_i\delta\varphi\partial^i\delta\varphi
    +{\cal A}_4a^{-4}\partial^2\delta\varphi\partial_i\delta\varphi\partial^i\delta\varphi\nonumber\\
 &~&+{\cal A}_5a^{-2}\partial^2\delta\varphi\partial_i\delta\varphi\partial^i\Psi+{\cal A}_6\dot{\delta\varphi}^2\delta\varphi
    +{\cal A}_7a^{-2}\dot{\delta\varphi}^2\partial^2\delta\varphi+{\cal A}_8a^{-2}\dot{\delta\varphi}\partial_i\delta\varphi\partial^i\dot{\delta\varphi}\nonumber\\
 &~&+{\cal A}_9\dot{\delta\varphi}\partial_i\delta\varphi\partial^i\psi+{\cal A}_{10}a^{-2}\partial^2\dot{\delta\varphi}\partial_i\delta\varphi\partial^i\psi
    +{\cal A}_{11}a^{-2}\partial^2\delta\varphi\partial_i\dot{\delta\varphi}\partial^i\psi
    +{\cal A}_{12}a^{-2}\delta\varphi\partial_i\dot{\delta\varphi}\partial^i\delta\varphi\nonumber\\
 &~&+{\cal A}_{13}a^{-2}\dot{\delta\varphi}\partial_i\delta\varphi\partial^i\delta\varphi
    +{\cal A}_{14}a^{-4}\partial^2\delta\varphi\partial_i\delta\varphi\partial^i\dot{\delta\varphi}
    +{\cal A}_{15}a^{-4}\partial^2\dot{\delta\varphi}\partial_i\delta\varphi\partial^i\delta\varphi\nonumber\\
 &~&+{\cal A}_{16}\dot{\delta\varphi}\partial_i\delta\varphi\partial^i\Psi+{\cal A}_{17}a^{-2}\partial^2\delta\varphi\partial_i\delta\varphi\partial^i\psi
    +{\cal A}_{18}a^{-2}\partial^2\dot{\delta\varphi}\partial_i\delta\varphi\partial^i\Psi
    +{\cal A}_{19}a^{-2}\partial^2\delta\varphi\partial_i\dot{\delta\varphi}\partial^i\Psi\Big]~,
\eea
\end{widetext}
where we define $\partial^2\psi\equiv\dot{\delta\varphi}$, $\partial^2\Psi\equiv\delta\varphi$, and
\begin{widetext}
\bea
\frac{{\cal A}_{1}}{(\epsilon_{\phi}+12)}&=&\frac{{\cal A}_{17}}{(\epsilon_{\phi}+3)}=-2\frac{{\cal A}_{13}}{(3\epsilon_{\phi}-5)}=2\Bigg(\frac{\xi}{M^{2}}\Bigg)^{\frac{3}{2}}\frac{\sqrt{y}H^{2}}{M_{p}^{2}}~,\nonumber\\
{\cal A}_{2}&=&{\cal A}_{5}={\cal A}_{16}/3H=2H{\cal A}_{18}=2H{\cal A}_{19}=-2\frac{\xi}{M^{2}}[\frac{\sqrt{\xi y}}{M}\frac{H^{3}}{M_{p}^{2}}(\frac{3}{2}\epsilon_{\phi}-9)-\frac{V_{,\varphi}}{2M_{p}^{2}}]~,\nonumber\\
{\cal A}_{3}&=&-\frac{\xi}{M^{2}}[\frac{\sqrt{\xi y}}{M}\frac{H^{3}}{M_{p}^{2}}(\frac{3}{2}\epsilon_{\phi}-24)-\frac{V_{,\varphi}}{2M_{p}^{2}}]~,~
{\cal A}_{4}=\frac{5}{8}{\cal A}_{7}=\frac{5}{4}H{\cal A}_{14}=\frac{5}{2}H{\cal A}_{15}=\frac{5}{2}\Bigg(\frac{\xi}{M^{2}}\Bigg)^{\frac{3}{2}}\frac{\sqrt{y}H}{M_{p}^{2}}~,\nonumber\\ \nonumber\\
{\cal A}_{6}&=&-\frac{\xi}{M^{2}}[\frac{\sqrt{\xi y}}{M}\frac{H^{3}}{M_{p}^{2}}(3\epsilon_{\phi}+9)-\frac{V_{,\varphi}}{M_{p}^{2}}]~,~{\cal A}_{8}=-5{\cal A}_{10}=-5{\cal A}_{11}=5{\cal A}_{9}/6H^{2}=5\Bigg(\frac{\xi}{M^{2}}\Bigg)^{\frac{3}{2}}\frac{\sqrt{y}H}{M_{p}^{2}}(\epsilon_{\phi}+3)~,\nonumber\\
{\cal A}_{12}&=&-\frac{\xi}{M^{2}}\frac{1}{H}[\frac{\sqrt{\xi y}}{M}\frac{H^{3}}{M_{p}^{2}}(\frac{3}{2}\epsilon_{\phi}+3)-\frac{V_{,\varphi}}{2M_{p}^{2}}]~.
\eea
\end{widetext}
The 3-point cross correlations are defined as:
\bea\label{correlationsss}
&&\langle\delta\varphi({\bf k}_1,t)\delta\varphi({\bf k}_2,t)\delta\varphi({\bf k}_3,t)\rangle\nonumber\\
&=&-i\langle|\int_{t_i}^{t_f}dt^\prime [\delta\varphi({\bf k}_1,t)\delta\varphi({\bf k}_2,t)\delta\varphi({\bf k}_3,t),\nonumber\\ &&H^{(3)}_{int}(t^\prime)]|\rangle~,
\eea
where for third order we have $H^{(3)}_{int}=-L^{(3)}_{sss}$. Here we choose $t_i$ to be infinite past, which corresponds to the Bunch-Davies vacuum, and $t_f$ to be some cutoff time scale, $t_c$ In inflationary scenario, $t_c$ should be reheating time, while for bounce scenario, $t_c$ corresponds to bouncing point. One can also use its conformal correspondence, namely $\eta_c$. In inflationary scenario it goes to $0$ and in bouncing scenario it is $\eta_B$. It is a very small number, when $\eta_c\rightarrow 0$, $\cos(K\eta_c)\rightarrow 1$, $\sin(K\eta_c)/(K\eta_c)\rightarrow 1$. We keep $\eta_c$ to avoid IR divergence in some of the following terms, which will be seen later. Because $\eta_c$ is small, we neglect higher order terms of $\eta_c$. The shape function is related to correlation function via the relation:
\bea\label{correlationsss2}
&&\langle\delta\varphi({\bf k}_1,t)\delta\varphi({\bf k}_2,t)\delta\varphi({\bf k}_3,t)\rangle\nonumber\\
&=&\frac{(2\pi)^3\delta^3(\sum_{i=1}^3{\bf k}_i)}{\Pi_{i=1}^3k_i^{3}}\frac{H^4}{M_p^4c_s^2Q_s^2}{\cal A}(k_1,k_2,k_3)~,
\eea
where we have used (\ref{correlation}), (\ref{shape}) and the power spectrum result (\ref{spectrum}).

Since there are two many terms in this part, we would like to classify these terms in terms of numbers of time derivatives, namely:
\subsubsection{parts of 3 time-derivatives}
This part only contains one term: ${\cal A}_1$, so the Hamiltonian is $H^{(3,3d)}_{int}=-\int d^3x a^3{\cal A}_1\dot{\delta\varphi}^3$. Substitute it into (\ref{correlationsss}) one can get the cross correlations of this part:
\bea
&&\langle\delta\varphi({\bf k}_1,t)\delta\varphi({\bf k}_2,t)\delta\varphi({\bf k}_3,t)\rangle^{(3d)}\nonumber\\
&\simeq&(2\pi)^3\delta^3(\sum_{i=1}^3{\bf k}_i){\cal A}_1\frac{3H^5}{8Q_s^3k_1^3k_2^3k_3^3}\frac{\mathbb{K}^6}{K^3}\cos(K\eta_c)~,
\eea
where we define $K\equiv k_1+k_2+k_3$, $\mathbb{K}^3\equiv k_1k_2k_3$. Comparing with (\ref{correlationsss2}) one gets the shape function of this part:
\be\label{A3d}
{\cal A}^{(3d)}(k_1,k_2,k_3)\simeq\frac{3Hc_s^2}{8Q_s}{\cal A}_1\frac{\mathbb{K}^6}{K^3}\cos(K\eta_c)~,
\ee

\begin{figure}[htbp]
\centering
\includegraphics[scale=0.3]{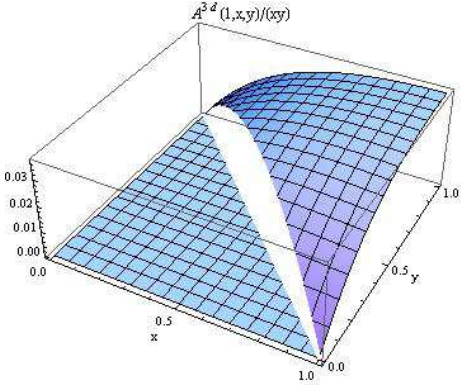}
\caption{The shape of the pure scalar part of the bispectrum which contains 3 time-derivatives in each term.  The shape peaks in the region where $x=y\rightarrow 1$, corresponding to an equilateral limit.}\label{shapeSSS3d}
\end{figure}

In Fig. \ref{shapeSSS3d} we plot the shape function ${\cal A}^{3d}$ in which the wavenumbers are normalized with $k_1=1$. We can see that, there is a peak in the region where $x=y\rightarrow 1$, namely $k_1\approx k_2\approx k_3$, which corresponds to an equilateral limit.

\subsubsection{parts of 2 time-derivatives}
This part contains six terms, namely ${\cal A}_6$, ${\cal A}_7$, ${\cal A}_8$, ${\cal A}_9$, ${\cal A}_{10}$, ${\cal A}_{11}$. Subsitute the Hamiltonian (which is the opposite of the Lagrangian) of these terms $H^{(3,2d)}_{int}$ into (\ref{correlationsss}) one can get the cross correlations of this part:
\bea
&&\langle\delta\varphi({\bf k}_{1},t)\delta\varphi({\bf k}_{2},t)\delta\varphi({\bf k}_{3},t)\rangle^{(2d)}~\nonumber\\
&\simeq&\frac{(2\pi)^{3}\delta^{3}(\sum_{i=1}^{3}{\bf k}_{i})H^{4}}{2^{5}Q_{s}^{3}k_{1}^{3}k_{2}^{3}k_{3}^{3}}\frac{k_{1}^{2}k_{2}^{2}}{K^{2}}\Bigg[(K+k_{3})\Big({\cal A}_{6}+{\cal A}_{9}\frac{{\bf k}_{2}\cdot{\bf k}_{3}}{k_{2}^{2}}\Big)~\nonumber\\
&&+\frac{2H^{2}(K+3k_{3})}{K^{2}}\Big({\cal A}_{7}k_{3}^{2}+{\cal A}_{8}({\bf k}_{2}\cdot{\bf k}_{3})+{\cal A}_{10}\frac{k_{1}^{2}({\bf k}_{2}\cdot{\bf k}_{3})}{k_{2}^{2}}\nonumber\\
&&+{\cal A}_{11}\frac{({\bf k}_{1}\cdot{\bf k}_{2})k_{3}^{2}}{k_{2}^{2}}\Big)\Bigg]\cos(K\eta_c)+5~perms.,
\eea
and comparing with (\ref{correlationsss2}) one gets the shape function of this part:
\bea\label{A2d}
&&{\cal A}^{(2d)}(k_{1},k_{2},k_{3})\nonumber\\
&\simeq&\frac{c_{s}^{2}}{2^{5}Q_{s}}\Bigg\{\frac{2{\cal A}_{6}}{K^{2}}\sum_{i\neq j}(k_{i}^{3}k_{j}^{2}+\mathbb{K}^{3}k_{i}k_{j})+\frac{{\cal A}_{9}}{2K^{2}}\Big[2\sum_{i=1}^{3}k_{i}^{5}\nonumber\\
&&+\sum_{i\neq j}(3k_{i}^{4}k_{j}-5k_{i}^{3}k_{j}^{2}-3\mathbb{K}^{3}k_{i}k_{j})\Big]\nonumber\\
&&+(2{\cal A}_{7}-{\cal A}_{8})\frac{12H^{2}\mathbb{K}^{6}}{K^{3}}+{\cal A}_{10}\frac{H^{2}}{K^{4}}\Big[2\sum_{i=1}^{3}k_{i}^{7}+\sum_{i\neq j}(5k_{i}^{6}k_{j}\nonumber\\
&&-2k_{i}^{5}k_{j}^{2}-5k_{i}^{4}k_{j}^{3}-5\mathbb{K}^{3}k_{i}^{3}k_{j})\Big]+{\cal A}_{11}\frac{3H^{2}}{K^{4}}\Big[\sum_{i\neq j}(k_{i}^{5}k_{j}^{2}\nonumber\\ &&-k_{i}^{4}k_{j}^{3})-4\mathbb{K}^{6}K\Big]\Bigg\}\cos(K\eta_c)~.
\eea

In the following we plot the shape functions of the bispectra given in the above results. Although there are totally six parts of bispectrum, there are less kinds of shapes since some parts actually give quite the similar shapes. Therefore in the following, we will only plot representative ones which are distinctive from each other, while contributions with the same shape will be addressed in their captions. The same way applies for other cases.

From the plots we can see that, the ${\cal A}_6$ part gives rise to shape function which peaks on its squeezed limit, while shapes of ${\cal A}_7$, ${\cal A}_8$, ${\cal A}_9$ and ${\cal A}_{11}$ peaks on their equilateral limit. Moreover, the ${\cal A}_{10}$ generates peaks on both enfolded and equilateral limit, showing an orthogonal feature.

\begin{figure}[htbp]
\centering
\includegraphics[scale=0.3]{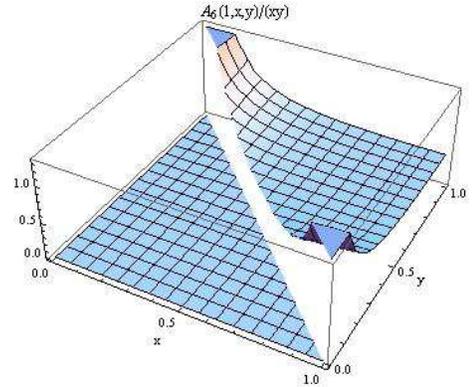}
\caption{The shape of the bispectrum of contribution ${\cal A}_6$. The shape peaks in the region where $x\rightarrow 1$, $y\rightarrow 0$ and vice versa, corresponding to a squeezed limit. }\label{shapeSSS2d6}
\end{figure}

\begin{figure}[htbp]
\centering
\includegraphics[scale=0.3]{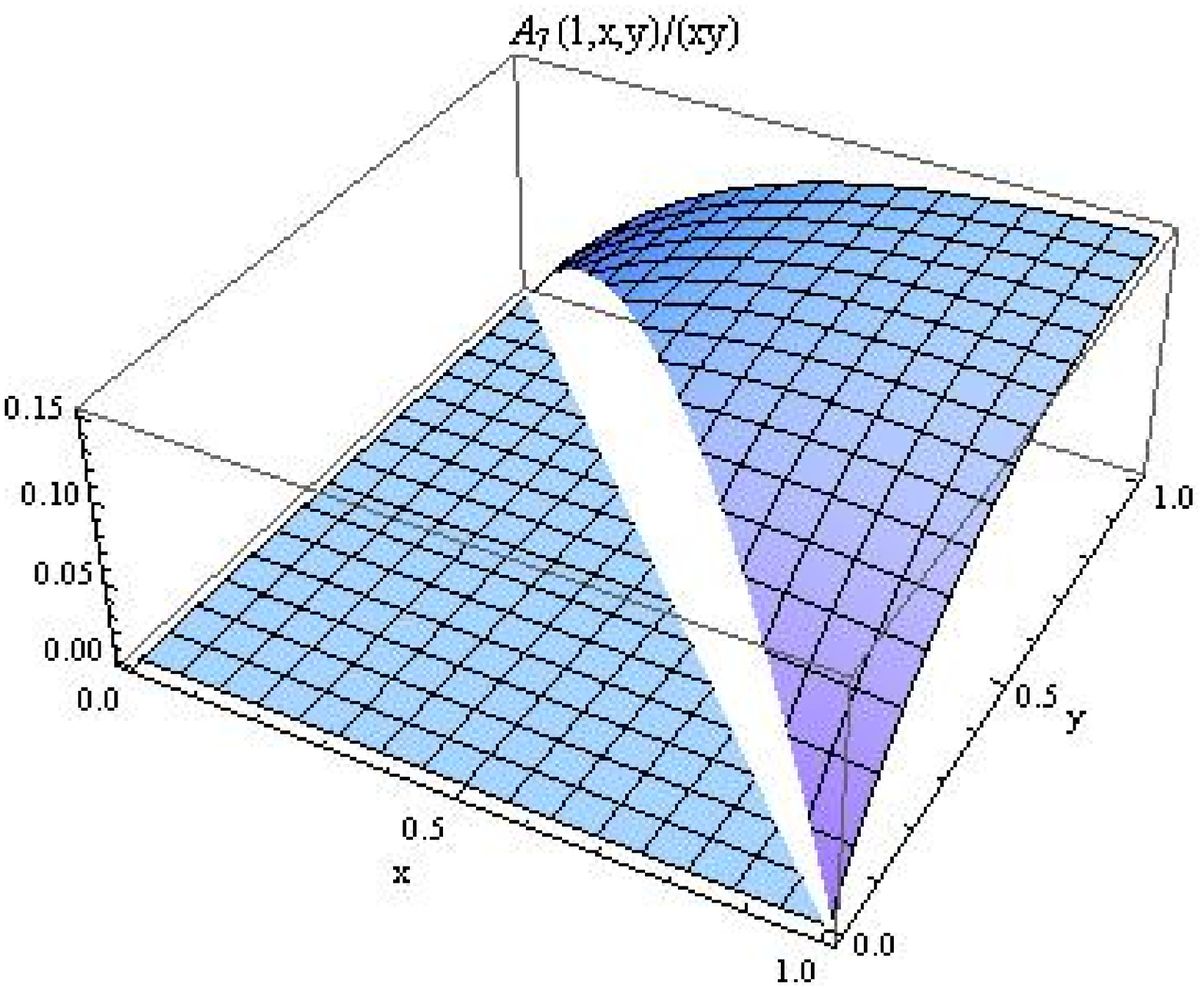}
\caption{The shape of the bispectrum of contribution ${\cal A}_7$. The shape peaks in the region where $x=y\rightarrow 1$, corresponding to an equilateral limit. The same shape are also given by contributions ${\cal A}_8$, ${\cal A}_9$ and ${\cal A}_{11}$.}\label{shapeSSS2d7}
\end{figure}

\begin{figure}[htbp]
\centering
\includegraphics[scale=0.3]{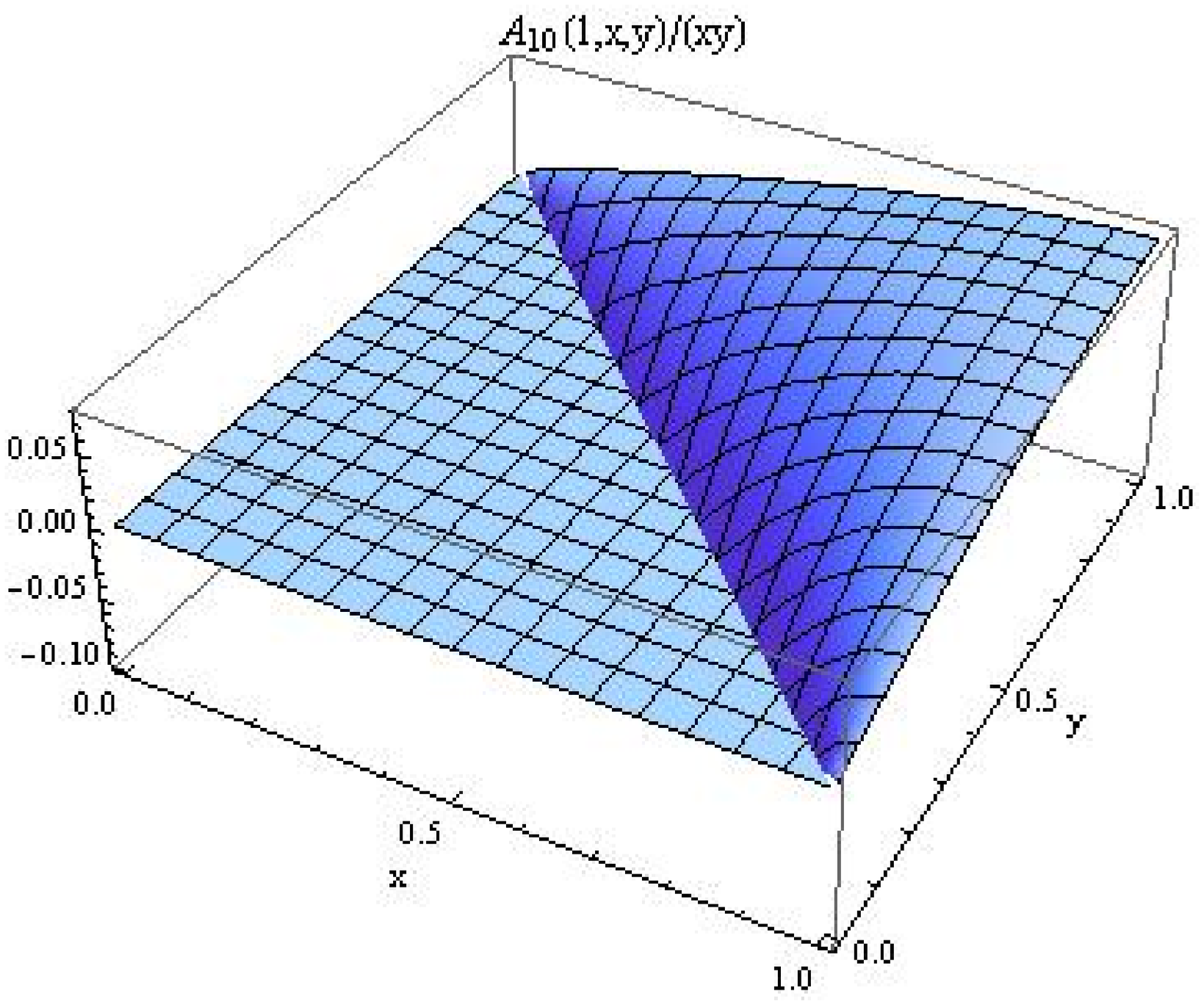}
\caption{The shape of the bispectrum of contribution ${\cal A}_{10}$. The shape peaks both in the region where $x=y\rightarrow 1$, corresponding to an equilateral limit, and in the region where $x+y=1$, corresponding to a folding limit.}\label{shapeSSS2d10}
\end{figure}

\subsubsection{parts of 1 time-derivative}
This part contains eight terms, namely ${\cal A}_{12}$, ${\cal A}_{13}$, ${\cal A}_{14}$, ${\cal A}_{15}$, ${\cal A}_{16}$, ${\cal A}_{17}$, ${\cal A}_{18}$, ${\cal A}_{19}$. Subsitute the Hamiltonian of these terms $H^{(3,1d)}_{int}$ into (\ref{correlationsss}) one can get the cross correlations of this part:
\bea
&&\langle\delta\varphi({\bf k}_{1},t)\delta\varphi({\bf k}_{2},t)\delta\varphi({\bf k}_{3},t)\rangle^{(1d)}\nonumber\\ &&=\frac{(2\pi)^{3}\delta^{3}(\sum_{i=1}^{3}{\bf k}_{i})H^{3}}{2^{5}Q_s^{3}k_{1}^{3}k_{2}^{3}k_{3}^{3}}\Bigg\{2{\cal A}_{16}k_{1}^{2}\frac{{\bf k}_{2}\cdot{\bf k}_{3}}{k_{3}^{2}}\Big(\frac{k_1k_3}{K}-\frac{\sin(K\eta_c)}{\eta_c}\nonumber\\
&&-k_2\Re[Ci(K\eta_c)]\Big)+k_{1}^{2}\Bigg[\frac{H^{4}}{K^{5}}({\cal A}_{14}({\bf k}_{1}\cdot{\bf k}_{2})k_{3}^{2}\nonumber\\
&&+{\cal A}_{15}k_{1}^{2}({\bf k}_{2}\cdot{\bf k}_{3}))\Big(2K^{2}+6\sum_{i=1}^3k_{i}^{2}+6\sum_{i\neq j}k_{i}k_{j}-6k_{1}^{2}\nonumber\\
&&+30k_{2}k_{3}\Big)+\frac{H^{2}}{K^{3}}\Big({\cal A}_{12}({\bf k}_{1}\cdot{\bf k}_{2})+{\cal A}_{13}({\bf k}_{2}\cdot{\bf k}_{3})\nonumber\\
&&+{\cal A}_{17}\frac{({\bf k}_{1}\cdot{\bf k}_{2})k_{3}^{2}}{k_{1}^{2}}+{\cal A}_{18}\frac{k_{1}^{2}({\bf k}_{2}\cdot{\bf k}_{3})}{k_{3}^{2}}+{\cal A}_{19}\frac{({\bf k}_{1}\cdot{\bf k}_{2})k_{3}^{2}}{k_{2}^{2}}\Big)\nonumber\\
&&\Big(K^{2}+\frac{1}{2}\sum_{i=1}^3k_{i}^{2}+\sum_{i\neq j}k_{i}k_{j}-k_{1}^{2}+3k_{2}k_{3}\Big)\Bigg]\cos(K\eta_c)\Bigg\}\nonumber\\
&&+5~perms.
\eea
Comparing with (\ref{correlationsss2}) one gets the shape function of this part:
\bea\label{A1d}
&&{\cal A}^{(1d)}(k_{1},k_{2},k_{3})\nonumber\\
&=&\frac{c_{s}^{2}}{2^{4}HQ_{s}}\frac{{\cal A}_{16}}{\mathbb{K}^6K}\Bigg\{\mathbb{K}^3\sum_{i\neq j}(k_{i}^{6}k_{j}-k_{i}^{4}k_{j}^{3})-\mathbb{K}^6\sum_{i\neq j}k_{i}^{3}k_{j}\nonumber\\
&&+[\sum_{i\neq j}(k_{i}^{5}k_{j}^5-k_{i}^{7}k_{j}^{3})-\mathbb{K}^3\sum_{i\neq j}(k_{i}^{5}k_{j}^2-k_{i}^{4}k_{j}^{3})\nonumber\\
&&+\mathbb{K}^6\sum_{i\neq j}(k_{i}^{3}k_{j}+k_{i}^{2}k_{j}^{2})+2\mathbb{K}^9K]\Re[Ci(K\eta_c)]\nonumber\\
&&+[-\sum_{i\neq j}(k_{i}^{7}k_{j}^2+k_{i}^{6}k_{j}^3-k_{i}^{5}k_{j}^{4})-\mathbb{K}^3\sum_{i\neq j}(k_{i}^{5}k_{j}-k_{i}^{3}k_{j}^{3})\nonumber\\
&&+2\mathbb{K}^6(\sum_{i=1}^3k_{i}^{3}+\sum_{i\neq j}k_{i}^{2}k_{j})]\frac{\sin(K\eta_c)}{\eta_c}\Bigg\}\nonumber\\
&&+\frac{M_{p}^{4}c_{s}^{2}}{2^{5}HQ_{s}}\Bigg\{{\cal A}_{14}\frac{3H^{4}}{K^{5}}\Big[\sum_{i\neq j}(k_{i}^{6}k_{j}^{2}\nonumber\\
&&-k_{i}^{4}k_{j}^{4})+5\mathbb{K}^{3}\sum_{i\neq j}(k_{i}^{4}k_{j}-k_{i}^{3}k_{j}^{2})-6\mathbb{K}^{6}\sum_{i=1}^{3}k_{i}^{2}\nonumber\\ &&-10\mathbb{K}^{6}\sum_{i\neq j}k_{i}k_{j}\Big]+{\cal A}_{15}\frac{2H^{4}}{K^{5}}\Big[\sum_{i=1}^{3}k_{i}^{8}+\sum_{i\neq j}(3k_{i}^{7}k_{j}\nonumber\\ &&+3k_{i}^{6}k_{j}^{2}-3k_{i}^{5}k_{j}^{3}-4k_{i}^{4}k_{j}^{4})+3\mathbb{K}^{3}(6\sum_{i=1}^{3}k_{i}^{5}-\sum_{i\neq j}k_{i}^{4}k_{j}\nonumber\\
&&-6\sum_{i\neq j}k_{i}^{3}k_{j}^{2})-8\mathbb{K}^{6}\sum_{i=1}^{3}k_{i}^{2}\Big]-{\cal A}_{12}\frac{H^{2}}{K^{3}}\Big[\sum_{i=1}^{3}k_{i}^{6}\nonumber\\
&&+\sum_{i\neq j}(k_{i}^{5}k_{j}+2k_{i}^{4}k_{j}^{2})+4\mathbb{K}^{3}\sum_{i=1}^{3}k_{i}^{3}\Big]+{\cal A}_{13}\frac{H^{2}}{K^{3}}\Big[\sum_{i=1}^{3}k_{i}^{6}\nonumber\\
&&+\sum_{i\neq j}(k_{i}^{5}k_{j}-k_{i}^{4}k_{j}^{2}-k_{i}^{3}k_{j}^{3})+4\mathbb{K}^{3}\sum_{i=1}^{3}k_{i}^{3}-5\mathbb{K}^{3}\sum_{i\neq j}k_{i}^{2}k_{j}\nonumber\\
&&-12\mathbb{K}^{6}\Big]+{\cal A}_{17}\frac{H^{2}}{2K^{3}}\Big[4\sum_{i=1}^{3}k_{i}^{6}+\sum_{i\neq j}(5k_{i}^{5}k_{j}-4k_{i}^{4}k_{j}^{2}\nonumber\\
&&-5k_{i}^{3}k_{j}^{3})+2\mathbb{K}^{3}\sum_{i=1}^{3}k_{i}^{3}-7\mathbb{K}^{3}\sum_{i\neq j}k_{i}^{2}k_{j}-18\mathbb{K}^{6}\Big]\nonumber\\
&&+{\cal A}_{18}\frac{H^{2}}{2\mathbb{K}^{6}K^{3}}\Big[\sum_{i\neq j}(k_{i}^{10}k_{j}^{2}+k_{i}^{9}k_{j}^{3}+k_{i}^{8}k_{j}^{4}-k_{i}^{7}k_{j}^{5}\nonumber\\
&&-2k_{i}^{6}k_{j}^{6})+\mathbb{K}^{3}\sum_{i\neq j}(k_{i}^{8}k_{j}+4k_{i}^{7}k_{j}^{2}-k_{i}^{6}k_{j}^{3}-4k_{i}^{5}k_{j}^{4})\nonumber\\ &&+2\mathbb{K}^{6}\sum_{i=1}^{3}k_{i}^{6}-2\mathbb{K}^{6}\sum_{i\neq j}(k_{i}^{5}k_{j}+3k_{i}^{4}k_{j}^{2})-8\mathbb{K}^{9}\sum_{i=1}^{3}k_{i}^{3}\Big]\nonumber\\
&&+{\cal A}_{19}\frac{H^{2}}{2\mathbb{K}^{6}K^{3}}\Big[\sum_{i\neq j}(k_{i}^{8}k_{j}^{4}-k_{i}^{6}k_{j}^{6})+3\mathbb{K}^{3}\sum_{i\neq j}(k_{i}^{6}k_{j}^{3}-k_{i}^{5}k_{j}^{4})\nonumber\\
&&-\mathbb{K}^{6}\sum_{i\neq j}(3k_{i}^{4}k_{j}^{2}+k_{i}^{3}k_{j}^{3})-5\mathbb{K}^{9}\sum_{i\neq j}k_{i}^{2}k_{j}-12\mathbb{K}^{12}\Big]\Bigg\}
\eea

From the plots we can see that the ${\cal A}_{12}$, ${\cal A}_{16}$ and ${\cal A}_{19}$ parts have shape functions which peak on their squeezed limits, and the ${\cal A}_{13}$, ${\cal A}_{14}$ and ${\cal A}_{17}$ parts have shape functions which peak on their equilateral limits. The ${\cal A}_{15}$ and ${\cal A}_{18}$ parts have different types of orthogonal shapes, which peaks on folded + equilateral and folded + squeezed limits, respectively. Moreover, since ${\cal A}_{16}$ will be divergent in the limit where $\eta_c\rightarrow 0$, it is useful to set a cutoff scale for ${\cal A}_{16}$, where we choose $K\eta_c=-0.001$. However, different choices of cutoff will hardly change our results.

\begin{figure}[htbp]
\centering
\includegraphics[scale=0.3]{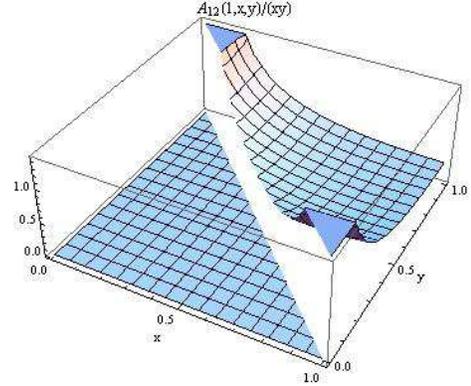}
\caption{The shape of the bispectrum of contribution ${\cal A}_{12}$. The shape peaks in the region where $x\rightarrow 1$, $y\rightarrow 0$ and vice versa, corresponding to a squeezed limit. The same shape are also given by contributions ${\cal A}_{16}$ and ${\cal A}_{19}$, where in ${\cal A}_{16}$ we choose the cut-off to be $K\eta_c=-0.001$.}\label{shapeSSS1d12}
\end{figure}

\begin{figure}[htbp]
\centering
\includegraphics[scale=0.3]{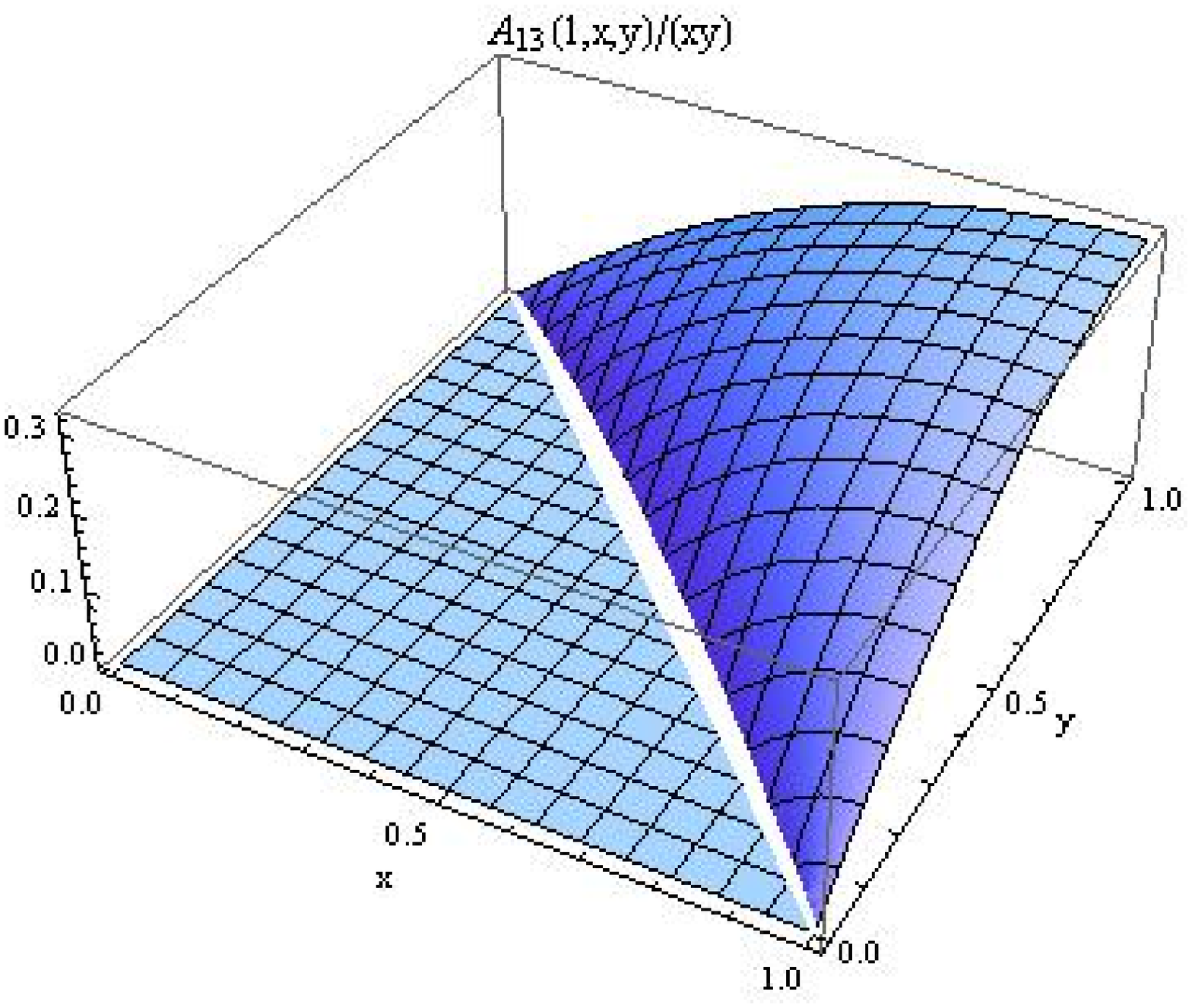}
\caption{The shape of the bispectrum of contribution ${\cal A}_{13}$. The shape peaks in the region where $x=y\rightarrow 1$, corresponding to an equilateral limit. The same shape are also given by contributions ${\cal A}_{14}$ and ${\cal A}_{17}$.}\label{shapeSSS1d13}
\end{figure}

\begin{figure}[htbp]
\centering
\includegraphics[scale=0.3]{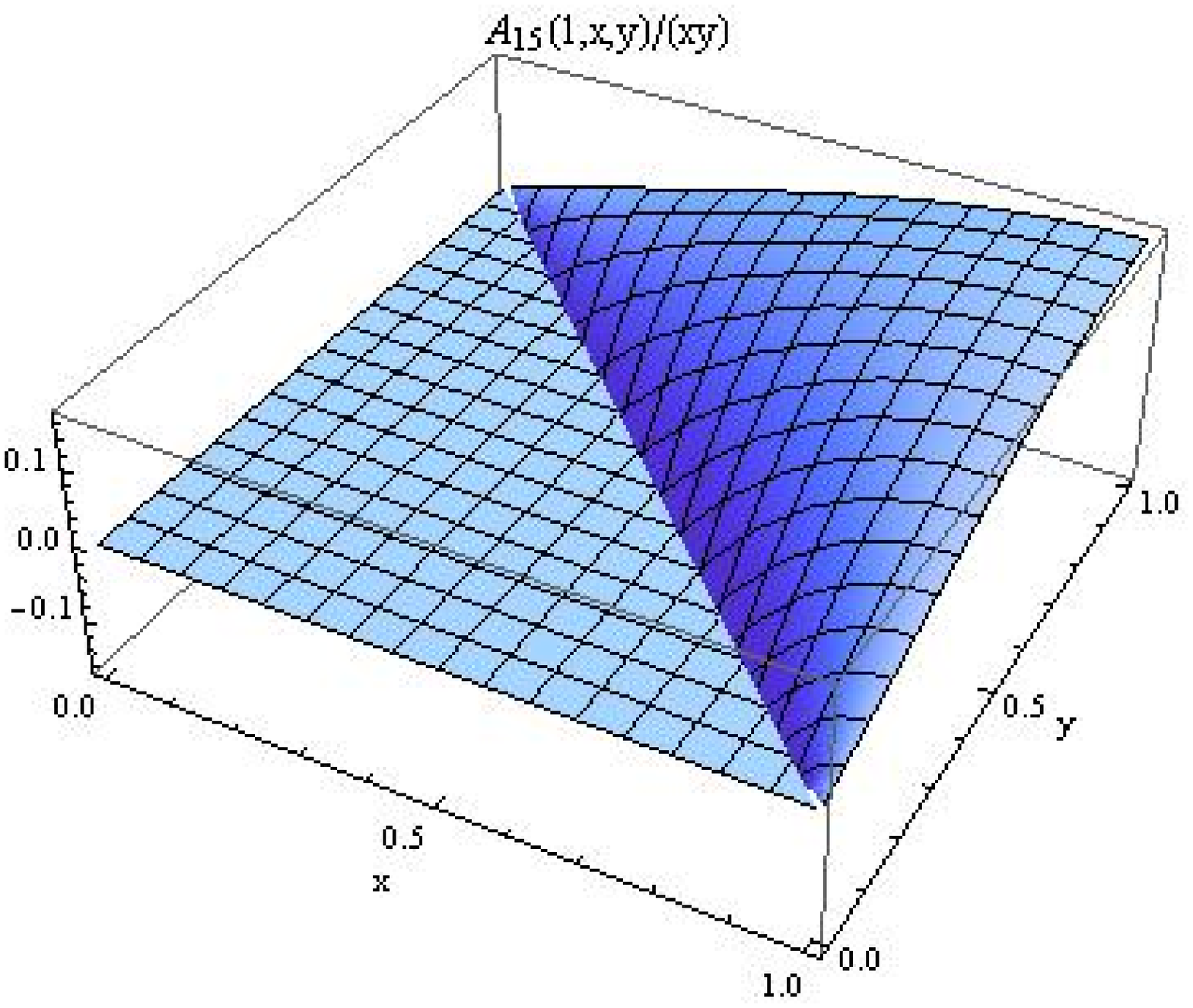}
\caption{The shape of the bispectrum of contribution ${\cal A}_{15}$. The shape peaks both in the region where $x=y\rightarrow 1$, corresponding to an equilateral limit, and in the region where $x+y=1$, corresponding to a folding limit.}\label{shapeSSS1d15}
\end{figure}

\begin{figure}[htbp]
\centering
\includegraphics[scale=0.3]{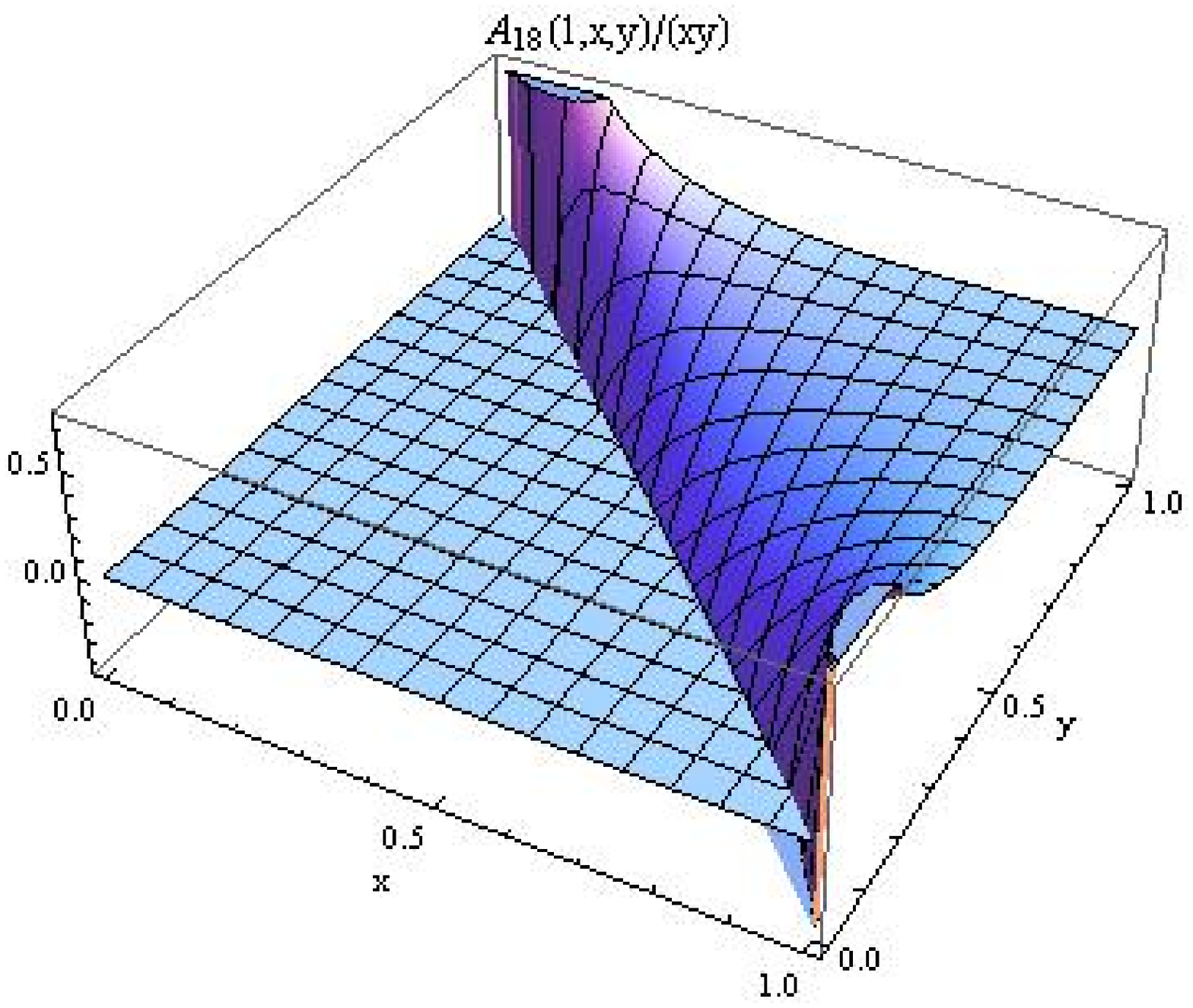}
\caption{The shape of the bispectrum of contribution ${\cal A}_{18}$. The shape peaks both in the region where $x\rightarrow 1$, $y\rightarrow 0$ and vice versa, corresponding to a squeezed limit, and in the region where $x+y=1$, corresponding to a folding limit.}\label{shapeSSS1d18}
\end{figure}

\subsubsection{parts of 0 time-derivative}
This part contains the last four terms, ${\cal A}_2$, ${\cal A}_3$, ${\cal A}_4$, ${\cal A}_5$. Subsitute the Hamiltonian of these terms $H^{(3,0d)}_{int}$ into (\ref{correlationsss}) one can get the cross correlations of this part:
\bea
&&\langle\delta\varphi({\bf k}_{1},t)\delta\varphi({\bf k}_{2},t)\delta\varphi({\bf k}_{3},t)\rangle^{(0d)}\nonumber\\ &\simeq&\frac{(2\pi)^{3}\delta^{3}(\sum_{i=1}^{3}{\bf k}_{i})H^{4}}{2^{5}Q_s^{3}k_{1}^{3}k_{2}^{3}k_{3}^{3}K^{2}}\Bigg\{{\cal A}_{4}H^{2}\frac{({\bf k}_{1}\cdot{\bf k}_{2})k_{3}^{2}}{K^{2}}\cos(K\eta_c)\Big[2k_{1}^{3}\nonumber\\
&&+8k_{1}^{2}(k_{2}+k_{3})+(6k_{1}+2K)(k_{2}^{2}+3k_{2}k_{3}+k_{3}^{2})\Big]\nonumber\\
&&+\Big[\Big(\sum_{i\neq j}k_i^2k_j+4\mathbb{K}^3\Big)\cos(K\eta_c)-K^2\frac{\sin(K\eta_c)}{\eta_c}\Big]\nonumber\\
&&\Big({\cal A}_{2}k_{1}^{2}+{\cal A}_{3}({\bf k}_{1}\cdot{\bf k}_{2})+{\cal A}_{5}\frac{({\bf k}_{1}\cdot{\bf k}_{2})k_{3}^{2}}{k_{2}^{2}}\Big)\Bigg\}+5~perms.
\eea
and comparing with (\ref{correlationsss2}) one gets the shape function of this part:
\bea\label{A0d}
&&{\cal A}^{(0d)}(k_{1},k_{2},k_{3})\nonumber\\
&=&\frac{c_{s}^{2}}{2^{5}Q_{s}K^{2}}\Bigg\{{\cal A}_{4}\frac{H^{2}}{K}\Big[2\sum_{i=1}^{3}k_{i}^{6}+2\sum_{i\neq j}(3k_{i}^{5}k_{j}-k_{i}^{4}k_{j}^{2}\nonumber\\
&&-3k_{i}^{3}k_{j}^{3})+6\mathbb{K}^{3}\Big(2\sum_{i=1}^{3}k_{i}^{3}-3\sum_{i\neq j}k_{i}^{2}k_{j}\Big)\Big]+\Big[(2{\cal A}_{2}-{\cal A}_{3})\nonumber\\
&&\big(\sum_{i=1}^{3}k_{i}^{2}\big)+\frac{{\cal A}_{5}}{2\mathbb{K}^6}\Big(\sum_{i\neq j}(k_i^6k_j^2-k_i^4k_j^4)-2\mathbb{K}^6\sum_{i=1}^{3}k_{i}^{2}\Big)\Big]\nonumber\\
&&\Big[\Big(\sum_{i\neq j}k_i^2k_j+4\mathbb{K}^3\Big)\cos(K\eta_c)-K^2\frac{\sin(K\eta_c)}{\eta_c}\Big]~.
\eea

From the plots we can see that the ${\cal A}_2$ and ${\cal A}_3$ part have shape functions which peak on its squeezed limit, the ${\cal A}_4$ part has shape function which peaks on its equilateral limit, while the ${\cal A}_5$ part has orthogonal shape function, which peaks both on its folded and squeezed limit.

\begin{figure}[htbp]
\centering
\includegraphics[scale=0.3]{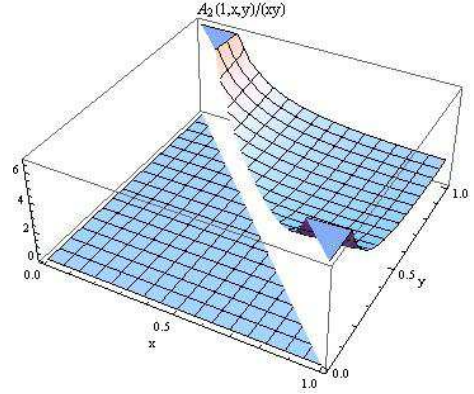}
\caption{The shape of the bispectrum of contribution ${\cal A}_2$. The shape peaks in the region where $x\rightarrow 1$, $y\rightarrow 0$ and vice versa, corresponding to a squeezed limit. The same shape are also given by contributions ${\cal A}_{3}$.}\label{shapeSSS0d2}
\end{figure}

\begin{figure}[htbp]
2\centering
\includegraphics[scale=0.3]{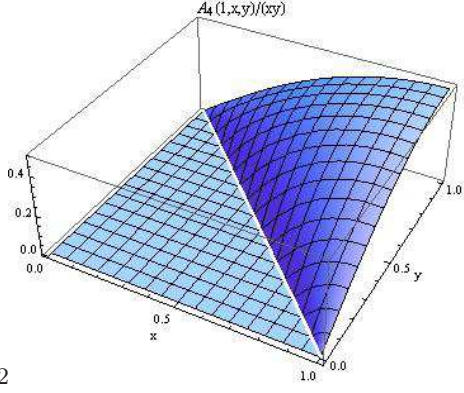}
\caption{The shape of the bispectrum of contribution ${\cal A}_4$. The shape peaks in the region where $x=y\rightarrow 1$, corresponding to an equilateral limit.}\label{shapeSSS0d4}
\end{figure}

\begin{figure}[htbp]
\centering
\includegraphics[scale=0.3]{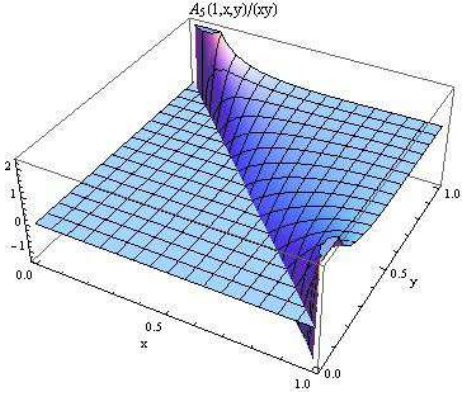}
\caption{The shape of the bispectrum of contribution ${\cal A}_{5}$. The shape peaks both in the region where $x\rightarrow 1$, $y\rightarrow 0$ and vice versa, corresponding to a squeezed limit, and in the region where $x+y=1$, corresponding to a folding limit.}\label{shapeSSS0d5}
\end{figure}

\subsubsection{The observables of Non-Gaussianity: $f_{NL}$}
After long derivations of the non-Gaussianities of scalar perturbation, let us now focus on the constraints on non-Gaussianities by observations, which is the most important and one of our main goals. People often use an estimator, $f_{NL}$, which is defined in Eq. (\ref{fnl}), to constraint non-Gaussianities. Although the general definition of $f_{NL}$ seems to be function of 3 $k_i$'s, there are three types of $f_{NL}$ of peculiar importance, which are:
\bea
f_{NL}^{eql}&=&f_{NL}|_{k_1\approx k_2\approx k_3}~\text{for equilateral type}~,\\
f_{NL}^{sqz}&=&f_{NL}|_{k_1\approx k_2,k_3\approx 0}~\text{for squeezed type}~,\\
f_{NL}^{enf}&=&f_{NL}|_{k_1\approx 2k_2\approx 2k_3}~\text{for enfolded type}~,
\eea
and the PLANCK data gives the very stringent constraints on equilateral and squeezed types of $f_{NL}$ as is shown before.

$f_{NL}$ relates to the shape functions ${\cal A}({\bf k}_1,{\bf k}_2,{\bf k}_3)$ as is given in Eq. (\ref{fnl}). In the above sections we have derived the shape functions ${\cal A}^{(3d)}$, ${\cal A}^{(2d)}$, ${\cal A}^{(1d)}$ and ${\cal A}^{(0d)}$ in Eqs. (\ref{A3d}), (\ref{A2d}), (\ref{A1d}) and (\ref{A0d}), respectively. However, these are only shape functions of dimentionful 3-point correlation functions, $\langle\delta\varphi\delta\varphi\delta\varphi\rangle$ while what we observed is that of dimensionless curvature perturbation, $\zeta$. Since according to $\delta N$ formalism we roughly have $\zeta\approx (H/\dot\varphi)\delta\varphi$ during inflation, one has ${\cal B}^{\langle\zeta\zeta\zeta\rangle}\approx(H/\dot\varphi)^3{\cal B}$ and ${\cal P}_\zeta\approx(H/\dot\varphi)^2{\cal P}_{\delta\varphi}$. From Eq. (\ref{shape}), the shape of $\langle\zeta\zeta\zeta\rangle$ is:
\be
{\cal A}^{\langle\zeta\zeta\zeta\rangle}\approx\left(\frac{\dot\varphi}{H}\right)({\cal A}^{(3d)}+{\cal A}^{(2d)}+{\cal A}^{(1d)}+{\cal A}^{(0d)})~.
\ee
Taking ${\cal A}^{\langle\zeta\zeta\zeta\rangle}$ back into Eq. (\ref{fnl}) and taking $k_1=k_2=k_3$ limits, we can get the equilateral type of $f_{NL}$:
\bea\label{fnleql}
f_{NL}^{eql}&=&\frac{5H^3\xi^{5/2}c_s^2M_p^2}{1296M^6y^{3/2}Q_s}\{2H^3(3572-637\epsilon_\phi)\sqrt{y\xi}\nonumber\\ &&+405MV_{,\varphi}-162\Re[Ci(3k\eta_c)][3H^3(6-\epsilon_\phi)\sqrt{y\xi}\nonumber\\
&&+MV_{,\varphi}]\}~,
\eea
where making use of Eq. (\ref{smally}) for $Q_s$ and $c_s^2$, taking the cut-off of $K\eta_c=-0.001$ and neglecting the potential term $V_{,\varphi}$ can make the formula greatly reduced. Moreover, for inflationary evolution where the slow-roll parameter $\epsilon_\phi$ is small, one can consider only the leading order of $\epsilon_\phi$, so the result will be:
\be\label{fnleql2}
f_{NL}^{eql}\simeq 10\frac{|y|}{M_p^2}(1+{\cal O}(\epsilon_\phi))~.
\ee
Moreover, as has been discussed in Sec. IIIB (see also earlier discussion in \cite{Feng:2013pba}), according to the instability requirement, our model can only allow for $|y|\ll M_p^2$. Under this requirement, one can expect a small $f_{NL}^{eql}$, which is well within the constraints of PLANCK data. For example, if we choose $|y|/M_p^2\sim 10^{-2}$, one can get:
\be
f_{NL}^{eql}\simeq 0.1~.
\ee

Furthermore, one can take different limits of $k_i$'s to get different types of $f_{NL}$, for instance, the squeezed and folded ones. Since the squeezed $f_{NL}$ is similar to the local type ones that has been obtained in \cite{Feng:2013pba}, we will not bother to recalculate it again in our present paper.
\subsection{Three-point correlation function: pure tensor part}
In this section, we calculate the non-Gaussianity of pure tensor part. The 3-rd order action of pure tensor part reads
\bea\label{actionttt}
S_{ttt}&\subset&\int dt L^{(3)}_{ttt}~\nonumber\\
&=&\frac{M_p^2}{4}\int dtd^3xa(\gamma^i_e\gamma^{ef}_{~,in}\gamma^n_f-\frac{1}{2}\gamma^{ij}\gamma^{mn}\gamma_{mn,ij})~,
\eea
where $Q_T$ has been given in (\ref{QTcT}). The pure tensor part of non-Gaussianity is given in Eq. (\ref{ng}) where for pure tensor part one can replace $\delta$ with $\gamma_{ij}$. Since there are no kinetic term in pure tensor part, one can identify the interacting Hamiltonian in (\ref{ng}) with its Lagrangian with a minus sign, namely,
\bea\label{correlationttt}
&&\langle\gamma_{i_1j_1}({\bf k}_1,t)\gamma_{i_2j_2}({\bf k}_2,t)\gamma_{i_3j_3}({\bf k}_3,t)\rangle\nonumber\\
&=&-i\langle|\int_{t_i}^{t_f}dt^\prime [\gamma_{i_1j_1}({\bf k}_1,t)\gamma_{i_2j_2}({\bf k}_2,t)\gamma_{i_3j_3}({\bf k}_3,t),\nonumber\\ &&H^{(3)}_{int}(t^\prime)]|\rangle~,
\eea
where $H^{(3)}_{int}=-L^{(3)}_{ttt}$. Substituting (\ref{fouriertensor}) and (\ref{tensorsub}) into (\ref{correlationttt}) and after long derivation, one gets:
\bea\label{correlationttt2}
&&\langle\gamma_{i_1j_1}({\bf k}_1,t)\gamma_{i_2j_2}({\bf k}_2,t)\gamma_{i_3j_3}({\bf k}_3,t)\rangle~\nonumber\\
&\simeq&\frac{(2\pi)^3H^4M_p^2}{8Q_T^3\Pi_{i=1}^3k_i^{3}}\delta^3(\sum_{i=1}^3{\bf k}_i)\Pi_{ij,i_2j_2}({\bf k}_2)\Pi_{mn,i_3j_3}({\bf k}_3)\nonumber\\
&&\times[\frac{1}{2}k_{1i}k_{1n}\Pi_{jm,i_1j_1}({\bf k}_1)-\frac{1}{4}k_{1i}k_{1j}\Pi_{mn,i_1j_1}({\bf k}_1)]\nonumber\\
&&\times\frac{1}{K^2}\Big[\Big(\sum_{i\neq j}k_i^2k_j+4\mathbb{K}^3\Big)\cos(K\eta_c)-K^2\frac{\sin(K\eta_c)}{\eta_c}\Big]\nonumber\\
&&+2~perms,
\eea
where $\Pi_{ij,kl}({\bf k})$ is defined as:
\be
\Pi_{ij,kl}({\bf k})\equiv\sum_{\lambda=1}^2e_{ij}({\bf k},\lambda)e_{kl}^\ast({\bf k},\lambda)~.
\ee

In order to compare with the usual definition of 3-point correlation function, (\ref{correlation}), we consider the non-indexed variable,
\be\label{redeftensor}
\gamma({\bf k},\lambda)=e_{ij}({\bf k},\lambda)\gamma_{ij}({\bf k})~,
\ee
then the non-indexed 3-point correlation functions can be obtained using (\ref{correlationttt2}):
\bea\label{correlationttt3}
&&\langle\gamma({\bf k}_1,\lambda_1)\gamma({\bf k}_2,\lambda_2)\gamma({\bf k}_3,\lambda_3)\rangle\nonumber\\
&\simeq&\frac{(2\pi)^3H^4M_p^2}{8Q_T^3\Pi_{i=1}^3k_i^{3}}\delta^3(\sum_{i=1}^3{\bf k}_i)e_{ij}({\bf k}_2,\lambda_2)e_{mn}({\bf k}_3,\lambda_3)\nonumber\\   &&\times[\frac{1}{2}k_{1i}k_{1n}e_{jm}({\bf k}_1,\lambda_1)-\frac{1}{4}k_{1i}k_{1j}e_{mn}({\bf k}_1,\lambda_1)]\nonumber\\
&&\times\frac{1}{K^2}\Big[\Big(\sum_{i\neq j}k_i^2k_j+4\mathbb{K}^3\Big)\cos(K\eta_c)-K^2\frac{\sin(K\eta_c)}{\eta_c}\Big]\nonumber\\
&&+2~perms.,\nonumber\\
&=&\frac{(2\pi)^3H^4M_p^2}{8Q_T^3\Pi_{i=1}^3k_i^{3}}\delta^3(\sum_{i=1}^3{\bf k}_i){\cal K}^{\lambda_1,\lambda_2,\lambda_3}(k_1,k_2,k_3)\nonumber\\
&&\times\frac{1}{K^2}\Big[\Big(\sum_{i\neq j}k_i^2k_j+4\mathbb{K}^3\Big)\cos(K\eta_c)-K^2\frac{\sin(K\eta_c)}{\eta_c}\Big]\nonumber\\
&&+2 perms.,
\eea
where
\begin{widetext}
\bea
&~&{\cal K}^{\lambda_1,\lambda_2,\lambda_3}(k_1,k_2,k_3)\nonumber\\
 &=&\frac{1}{128\sqrt{2}k_1^2k_2^2k_3^2}(k_1-k_2-k_3)(k_1+k_2-k_3)(k_1-k_2+k_3)(k_1+k_2+k_3)\nonumber\\
 &~&[3k_1^4-(k_2^2-k_3^2)^2+k_1k_2(k_3^2-k_2^2)\lambda_1\lambda_2+k_1^3\lambda_1(k_2\lambda_2+2k_3\lambda_3)+k_1^2(-2k_2^2+6k_3^2+2\lambda_2\lambda_3k_2k_3)]~,
\eea
\end{widetext}
Here we also used the relations (\ref{poltensor}). From Eqs. (\ref{correlation}) and (\ref{shape}), we define the shape function of the pure tensor part through the relation
\bea
&&\langle\gamma({\bf k}_1,\lambda_1)\gamma({\bf k}_2,\lambda_2)\gamma({\bf k}_3,\lambda_3)\rangle\nonumber\\
&=&\frac{(2\pi)^7\delta^3(\sum_{i=1}^3{\bf k}_i)}{\Pi_{i=1}^3k_i^{3}}{\cal P}_{\delta\varphi}^2{\cal A}^{\lambda_1,\lambda_2,\lambda_3}({\bf k}_1,{\bf k}_2,{\bf k}_3)~,
\eea
which gives
\bea
&&{\cal A}^{\lambda_1,\lambda_2,\lambda_3}({\bf k}_1,{\bf k}_2,{\bf k}_3)\simeq\frac{M_p^2Q_s^2c_s^2}{8Q_T^3}{\cal K}^{\lambda_1,\lambda_2,\lambda_3}(k_1,k_2,k_3)\nonumber\\
&&\times\frac{1}{K^2}\Big[\Big(\sum_{i\neq j}k_i^2k_j+4\mathbb{K}^3\Big)\cos(K\eta_c)-K^2\frac{\sin(K\eta_c)}{\eta_c}\Big]\nonumber\\
&&+2~perms.
\eea
Here ${\cal P}_{\delta\varphi}$ has been given in Eq. (\ref{spectrum}). There are actually $2^3=8$ shapes, according to $\lambda_i(i=1,2,3)$ being positive or negative. However, due to the symmetry, there are actually only one independent shape. In this context, we will only plot one of the shapes for illustration, while other shapes can be related.

\begin{figure}[htbp]
\centering
\includegraphics[scale=0.3]{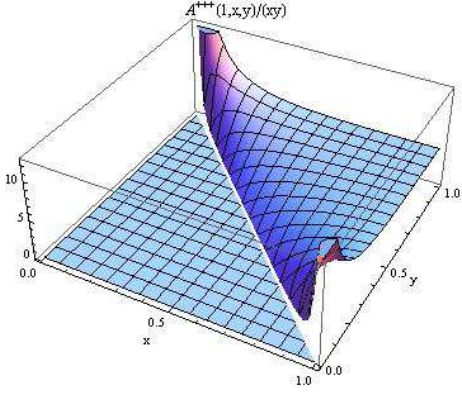}
\caption{The shape of the pure tensor part of the bispectrum with $\lambda_1=\lambda_2=\lambda_3=+2$. The shape peaks in the region where $x\rightarrow 1$, $y\rightarrow 0$ and vice versa, corresponding to a squeezed limit. }\label{shapeTTT}
\end{figure}

In Fig. \ref{shapeTTT} we plot the shape function ${\cal A}^{+++}$ which we choose $\lambda_1=\lambda_2=\lambda_3=+2$, while the wavenumbers are normalized with $k_1=1$. We can see that, there is a peak in the region where $x\rightarrow 1$, $y\rightarrow 0$, namely $k_1\approx k_2\gg k_3$, which corresponds to a squeezed limit.

\subsection{Three-point correlation function: 1 scalar+2 tensors}
In this section, we calculate the non-Gaussianity of the mixing parts which contain 1 scalar modes and 2 tensor mode. The 3-rd order action of 1 scalar+2 tensor part reads
\bea\label{actionsst}
S_{stt}&\subset&\int dt L^{(3)}_{stt}~\nonumber\\
&=&\int dtd^3x a^3[{\cal B}_1a^{-2}\dot{\delta\varphi}\gamma_{n,i}^m\gamma_{m,i}^n+{\cal B}_2a^{-2}\delta\varphi\gamma_{n,i}^m\gamma_{m,i}^n\nonumber\\
&&+{\cal B}_3\dot{\delta\varphi}\dot{\gamma}_{ij}\dot{\gamma}^{ij}+{\cal B}_4\delta\varphi\dot{\gamma}_{ij}\dot{\gamma}^{ij}+{\cal B}_5\partial^n\psi\dot{\gamma}_{ij}\gamma^{ij}_{~~,n}\nonumber\\
&&+{\cal B}_6\partial^n\Psi\dot{\gamma}_{ij}\gamma^{ij}_{~~,n}+{\cal B}_7a^{-2}\dot{\gamma}^k_j\gamma^j_{l,k}\partial^l\delta\varphi\nonumber\\
&&+{\cal B}_8a^{-2}\dot{\gamma}^k_j\gamma_{kl,j}\partial^l\delta\varphi]
\eea
where
\bea
{\cal B}_1&=&{\cal B}_3=-\frac{1}{2}{\cal B}_7=\frac{1}{2}{\cal B}_8=\frac{\xi\dot\varphi}{4M^2}~,\nonumber\\
{\cal B}_2&=&{\cal B}_4=-\frac{3}{8}\frac{\xi\dot\varphi H}{M^2}~,{\cal B}_5=\frac{\xi\dot\varphi H}{4M^2}(\epsilon_\phi+3)~,\nonumber\\
{\cal B}_6&=&-\frac{1}{4}[\frac{\xi\dot\varphi H^2}{M^2}(\frac{3}{2}\epsilon_\phi-9)-\frac{V_{,\varphi}}{2H}]~.
\eea
and the 3-point cross correlations are defined as:
\bea\label{correlationstt}
&&\langle\delta\varphi({\bf k}_1,t)\gamma_{i_2j_2}({\bf k}_2,t)\gamma_{i_3j_3}({\bf k}_3,t)\rangle\nonumber\\
&=&-i\langle|\int_{t_i}^{t_f}dt^\prime[\delta\varphi({\bf k}_1,t)\gamma_{i_2j_2}({\bf k}_2,t)\gamma_{i_3j_3}({\bf k}_3,t),\nonumber\\ &&H^{(3)}_{int}(t^\prime)]|\rangle~,
\eea
where in this case $H^{(3)}_{int}=-L^{(3)}_{stt}$. Substituting (\ref{actionstt}) into (\ref{correlationstt}) one can get:
\begin{widetext}
\bea\label{correlationstt2}
&&\langle\delta\varphi({\bf k}_1,t)\gamma_{i_2j_2}({\bf k}_2,t)\gamma_{i_3j_3}({\bf k}_3,t)\rangle\nonumber\\
 &=&\frac{H^6(\Pi_{i=1}^3k_i^{-3})}{8Q_T^2c_T^2Q_sc_s}(2\pi)^3\delta^3(\sum_{i=1}^3{\bf k}_i)\Big(\Pi_{ij,i_{2}j_{2}}({\bf k}_{2})\Pi_{ij,i_{3}j_{3}}({\bf k}_{3})\sum_{s=1}^6{\cal I}^{(s)}\Big)+\Pi_{ij,i_{2}j_{2}}({\bf k}_{2})\Pi_{jk,i_{3}j_{3}}({\bf k}_{3})(k_{1k}k_{3j})({\cal I}^{(7)}+{\cal I}^{(8)})\nonumber\\
 &&+(k_2\leftrightarrow k_3, i_2j_2\leftrightarrow i_3j_3)~,
\eea
\end{widetext}
where
\bea
{\cal I}^{(1)}&=&2c_{T}^{2}\frac{{\cal B}_{1}}{H}({\bf k}_{2}\cdot{\bf k}_{3})[c_{s}^{4}k_{1}^{4}+3c_{s}^{3}c_{T}k_{1}^{3}(k_{2}+k_{3})\nonumber\\ &&+2c_{T}^{2}c_{s}^{2}k_{1}^{2}(k_{2}^{2}+3k_{2}k_{3}+k_{3}^{2})]\cos[(c_sk_1+c_T(k_2\nonumber\\
&&+k_3))\eta_c]/[c_{s}k_{1}+c_{T}(k_{2}+k_{3})]^{3}~,\nonumber\\
{\cal I}^{(2)}&=&2c_{T}^{2}\frac{{\cal B}_{2}}{H^{2}}({\bf k}_{2}\cdot{\bf k}_{3})\{[4c_sc_{T}^{2}k_{1}k_{2}k_{3}+c_T^2c_{s}k_{1}^{2}(k_{2}+k_{3})\nonumber\\
&&+c_sc_{T}^{2}k_{1}(k_{2}^2+k_{3}^2)+c_{T}^{3}k_{2}k_{3}(k_{2}+k_{3})]\cos[(c_sk_1+c_T(k_2\nonumber\\
&&+k_3))\eta_c]-(c_sk_{1}+c_Tk_{2}+c_Tk_{3})^3\sin[(c_sk_1+c_T(k_2\nonumber\\
&&+k_3))\eta_c]/\eta_c\}/[c_{s}k_{1}+c_{T}(k_{2}+k_{3})]^{2}~,\nonumber\\
{\cal I}^{(3)}&=&4\frac{{\cal B}_{3}}{H}\frac{c_{s}^{2}c_{T}^{4}k_{1}^{2}k_{2}^{2}k_{3}^{2}\cos[(c_sk_1+c_T(k_2+k_3))\eta_c]}{[c_{s}k_{1}+c_{T}(k_{2}+k_{3})]^{3}}~,\nonumber\\
{\cal I}^{(4)}&=&2\frac{{\cal B}_{4}}{H^{2}}\frac{c_{T}^{4}k_{2}^{2}k_{3}^{2}[2c_{s}k_{1}+c_{T}(k_{2}+k_{3})]}{[c_{s}k_{1}+c_{T}(k_{2}+k_{3})]^{2}}\cos[(c_sk_1+c_T(k_2\nonumber\\
&&+k_3))\eta_c]~,\nonumber\\
{\cal I}^{(5)}&=&\frac{{\cal B}_{5}}{H^{2}}({\bf k}_{1}\cdot{\bf k}_{3})\frac{c_{s}c_{T}^{3}k_{2}^{2}[c_{s}k_{1}+c_{T}(k_{2}+2k_{3})]}{[c_{s}k_{1}+c_{T}(k_{2}+k_{3})]^{2}}\cos[(c_sk_1+c_T(k_2\nonumber\\
&&+k_3))\eta_c]~,\nonumber\\
\eea

\bea
{\cal I}^{(6)}&=&\frac{{\cal B}_{6}}{2H^{3}}\frac{c_{T}^{3}({\bf k}_{1}\cdot{\bf k}_{3})k_{2}^{2}}{c_{s}k_{1}^{2}}[\frac{\sin((c_{s}k_{1}+c_{T}k_{2}+c_{T}k_{3})\eta_c)}{\eta_c}\nonumber\\
&&-\frac{c_sc_T k_1k_3}{(c_{s}k_{1}+c_{T}k_{2}+c_{T}k_{3})}+\nonumber\\
&&c_Tk_2\Re[Ci((c_{s}k_{1}+c_{T}k_{2}+c_{T}k_{3})\eta_c)]]~,\nonumber\\
{\cal I}^{(7)}&=&+2c_{s}c_{T}^{3}\frac{{\cal B}_{7}}{H}[2c_{s}^{2}k_{1}^{2}+c_{T}(3c_{s}k_{1}+c_{T}k_{2}+c_{T}k_{3})\nonumber\\
&&(k_{2}+2k_{3})]k_{2}^{2}\cos[(c_sk_1+c_T(k_2\nonumber\\
&&+k_3))\eta_c]/[c_{s}k_{1}+c_{T}(k_{2}+k_{3})]^{3}~,\nonumber\\
{\cal I}^{(8)}&=&+2c_{s}c_{T}^{3}\frac{{\cal B}_{8}}{H}[2c_{s}^{2}k_{1}^{2}+c_{T}(3c_{s}k_{1}+c_{T}k_{2}+c_{T}k_{3})\nonumber\\
&&(k_{2}+2k_{3})]k_{2}^{2}\cos[(c_sk_1+c_T(k_2\nonumber\\
&&+k_3))\eta_c]/[c_{s}k_{1}+c_{T}(k_{2}+k_{3})]^{3}~.
\eea
However, note that since ${\cal B}_{7}=-{\cal B}_{8}$, one has ${\cal I}^{(7)}+{\cal I}^{(8)}=0$, so we only need to consider ${\cal I}^{(1)}\sim{\cal I}^{(6)}$. Making use of the redefinition of tensor mode (\ref{redeftensor}), one has:
\bea\label{correlationstt3}
&&\langle\delta\varphi({\bf k}_1)\gamma({\bf k}_2,\lambda_2)\gamma({\bf k}_3,\lambda_3)\rangle\nonumber\\
&=&\frac{(2\pi)^7\delta^3(\sum_{i=1}^3{\bf k}_i)}{(\Pi_{i=1}^3k_i^{3})}{\cal P}_{\delta\varphi}^2\sum_{s=1}^6{\cal A}^{\lambda_2,\lambda_3,s}({\bf k}_1,{\bf k}_2,{\bf k}_3)~,
\eea
which gives
\bea
&&{\cal A}^{\lambda_2,\lambda_3,s}({\bf k}_1,{\bf k}_2,{\bf k}_3)=\frac{H^2Q_sc_s}{8Q_T^2c_T^2}{\cal I}^{(s)}\nonumber\\
&&\times e_{i_2j_2}({\bf k}_2,\lambda_2)e_{i_3j_3}({\bf k}_3,\lambda_3)\Pi_{ij,i_{2}j_{2}}({\bf k}_{2})\Pi_{ij,i_{3}j_{3}}({\bf k}_{3})~.
\eea

There are actually $2^2=4$ shapes, according to $\lambda_i(i=1,2)$ being positive or negative. However, due to the symmetry, there are actually only one independent shape. In this context, we will only plot one of the shapes for illustration, while other shapes can be related.

\begin{figure}[htbp]
\centering
\includegraphics[scale=0.3]{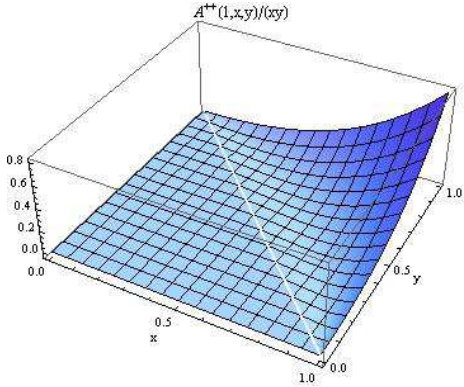}
\caption{The shape of the 1 scalar+2 tensors part of the bispectrum with $\lambda_1=\lambda_2=+2$. The shape peaks in the region where $x=y\rightarrow 1$, corresponding to an equilateral limit. For $B_6$, we choose the cut-off to be $(c_{s}k_{1}+c_{T}k_{2}+c_{T}k_{3})\eta_c=0.001$.}\label{shapeSTT}
\end{figure}

In Fig. \ref{shapeSTT} we plot the shape function ${\cal A}^{++}$ which we choose $\lambda_1=\lambda_2=+2$, while the wavenumbers are normalized with $k_1=1$. We can see that, there is a peak in the region where $x=y\rightarrow 1$, namely $k_1\approx k_2\approx k_3$, which corresponds to an equilateral limit.

\subsection{Three-point correlation function: 2 scalars+1 tensor}
As the last part, let us move on to the case of the non-Gaussianity of the mixing parts which contain 2 scalar modes and 1 tensor mode. The 3-rd order action of 2 scalar+1 tensor part reads
\bea\label{actiontss}
S_{tss}&\subset&\int dt L^{(3)}_{tss}~\nonumber\\
&=&\int dtd^3xa\big[{\cal C}_1\dot{\gamma}^{ij}\partial_i\dot{\delta\varphi}\partial_j\delta\varphi\nonumber\\
&&+{\cal C}_2\gamma^{ij}\partial_i\dot{\delta\varphi}\partial_j\delta\varphi+a^{-1}{\cal C}_3\gamma^{ij}_{~~,nn}\partial_i\delta\varphi\partial_j\delta\varphi\nonumber\\
&&+{\cal C}_4\dot{\gamma}^{ij}\partial_i\delta\varphi\partial_j\delta\varphi+{\cal C}_5\gamma^{ij}\partial_i\delta\varphi\partial_j\delta\varphi\big]~,
\eea
where
\be
{\cal C}_1=2{\cal C}_3=-\frac{\xi}{M^2},~{\cal C}_2=\frac{4}{5}{\cal C}_4=-\frac{4\xi H}{M^2},~{\cal C}_5=\frac{5\xi H^2}{M^2}~.
\ee
and the 3-point cross correlations are defined as:
\bea\label{correlationtss}
&&\langle\gamma_{ij}({\bf k}_1,t)\delta\varphi({\bf k}_2,t)\delta\varphi({\bf k}_3,t)\rangle\nonumber\\
&=&-i\langle|\int_{t_i}^{t_f}dt^\prime [\gamma_{ij}({\bf k}_1,t)\delta\varphi({\bf k}_2,t)\delta\varphi({\bf k}_3,t),\nonumber\\ &&H^{(3)}_{int}(t^\prime)]|\rangle~,
\eea
where in this case $H^{(3)}_{int}=-L^{(3)}_{tss}$. Substituting (\ref{actiontss}) into (\ref{correlationtss}) one can get:
\bea\label{correlationtss2}
&&\langle\gamma_{ij}({\bf k}_1)\delta\varphi({\bf k}_2)\delta\varphi({\bf k}_3)\rangle\nonumber\\
 &=&\frac{H^6(\Pi_{i=1}^3k_i^{-3})}{4Q_Tc_TQ_s^2c_s^2}(2\pi)^3\delta^3(\sum_{i=1}^3{\bf k}_i)\Pi_{i^\prime j^\prime,ij}({\bf k}_{1})(k_{2i^\prime}k_{3j^\prime})\nonumber\\
&&\sum_{s=1}^5{\cal J}^{(s)}+(k_2\leftrightarrow k_3)~,
\eea
where
\bea
{\cal J}^{(1)}&=&2{\cal C}_{1}c_{s}^{2}c_T^2k_{1}^{2}k_{2}^{2}\frac{(c_T k_{1}+c_{s}k_{2}+4c_{s}k_{3})}{(c_T k_{1}+c_{s}k_{2}+c_{s}k_{3})^{4}}\cos[(c_Tk_1\nonumber\\
&&+c_s(k_2+k_3))\eta_c]\nonumber\\
{\cal J}^{(2)}&=&\frac{{\cal C}_{2}}{H}c_{s}^{2}k_{2}^{2}[c_{s}(3c_T k_{1}+c_{s}k_{2}+c_{s}k_{3})(k_{2}+2k_{3})\nonumber\\
&&+2c_T^2k_{1}^{2}]\cos[(c_Tk_1+c_s(k_2\nonumber\\
&&+k_3))\eta_c]/(c_T k_{1}+c_{s}k_{2}+c_{s}k_{3})^{3}\nonumber\\
{\cal J}^{(3)}&=&{\cal C}_{3}c_T^2k_{1}^{2}[2c_T^3k_{1}^{3}+8c_T^2 c_{s}k_{1}^{2}(k_{2}+k_{3})\nonumber\\
&&+c_{s}^{2}(8c_T k_{1}+2c_{s}k_{2}+2c_{s}k_{3})(k_{2}^{2}+3k_{2}k_{3}\nonumber\\
&&+k_{3}^{2})]\cos[(c_Tk_1+c_s(k_2\nonumber\\
&&+k_3))\eta_c]/(c_T k_{1}+c_{s}k_{2}+c_{s}k_{3})^{4}\nonumber\\
{\cal J}^{(4)}&=&\frac{{\cal C}_{4}}{H}[c_T^4k_{1}^{4}+2c_{s}^{2}c_T^2k_{1}^{2}(k_{2}^{2}+3k_{2}k_{3}+k_{3}^{2})\nonumber\\
&&+3c_T^3 c_{s}k_{1}^3(k_{2}+k_{3})]\cos[(c_Tk_1+c_s(k_2\nonumber\\
&&+k_3))\eta_c]/(c_T k_{1}+c_{s}k_{2}+c_{s}k_{3})^{3}\nonumber\\
{\cal J}^{(5)}&=&\frac{{\cal C}_{5}}{H^{2}}\{[c_T^2c_{s}k_{1}^{2}(k_{2}+k_{3})+c_Tc_{s}^{2}k_{1}(k_{2}^2+k_{3}^2)\nonumber\\
&&+c_{s}^{3}k_{2}k_{3}(k_{2}+k_{3})+4c_Tc_{s}^{2}k_{1}k_{2}k_{3}]\cos[(c_Tk_1+c_s(k_2\nonumber\\
&&+k_3))\eta_c]-(c_Tk_{1}+c_sk_{2}+c_sk_{3})^3\sin[(c_Tk_1+c_s(k_2\nonumber\\
&&+k_3))\eta_c]/\eta_c\}/(c_T k_{1}+c_{s}k_{2}+c_{s}k_{3})^{2}
\eea
and making use of the redefinition of tensor mode (\ref{redeftensor}), one has:
\bea\label{correlationtss3}
&&\langle\gamma({\bf k}_1,\lambda)\delta\varphi({\bf k}_2)\delta\varphi({\bf k}_3)\rangle\nonumber\\
&=&\frac{(2\pi)^7\delta^3(\sum_{i=1}^3{\bf k}_i)}{(\Pi_{i=1}^3k_i^{3})}{\cal P}_{\delta\varphi}^2\sum_{s=1}^5{\cal A}^{\lambda,s}({\bf k}_1,{\bf k}_2,{\bf k}_3)~,
\eea
which gives
\bea
&&{\cal A}^{\lambda,s}({\bf k}_1,{\bf k}_2,{\bf k}_3)\nonumber\\
&=&\frac{H^2}{4Q_Tc_T}e_{ij}({\bf k}_{1},\lambda)\Pi_{i^\prime j^\prime,ij}({\bf k}_{1})(k_{2i^\prime}k_{3j^\prime}){\cal J}^{(s)}~.
\eea

There are two shapes, according to $\lambda$ being positive or negative. However, due to the symmetry, there are actually only one independent shape. In this context, we will only plot one of the shapes for illustration, while other shapes can be related.

\begin{figure}[htbp]
\centering
\includegraphics[scale=0.3]{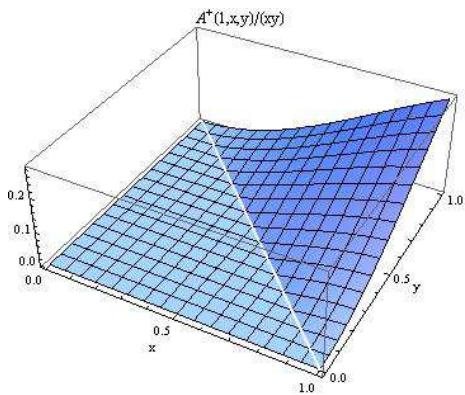}
\caption{The shape of the 2 scalars+1 tensor part of the bispectrum with $\lambda=+2$. The shape peaks in the region where $x=y\rightarrow 1$, corresponding to an equilateral limit. }\label{shapeSST}
\end{figure}

In Fig. \ref{shapeSST} we plot the shape function ${\cal A}^{+}$ which we choose $\lambda=+2$, while the wavenumbers are normalized with $k_1=1$. We can see that, there is a peak in the region where $x=y\rightarrow 1$, namely $k_1\approx k_2\approx k_3$, which corresponds to an equilateral limit.

From the analysis on non-Gaussianities of the pure tensor and mixed parts we can see that, unlike the pure scalar part, they can give rise to shape functions very concordantly peaking on squeezed and equilateral limit. This means that, the pure tensor and mixed parts of the perturbations will generate relatively large $f_{NL}^{sqz}$ and $f_{NL}^{eql}$, which can be tested by future observations. If we could find sizable non-Gaussianities of $f_{NL}^{sqz}$ for pure tensor and $f_{NL}^{eql}$ for mixed parts in the future, it will be a good support to our model.

\section{conclusion}
In this paper, we study the full description of the model of curvaton with nonminimal derivative coupling to Einstein Gravity, up to 3rd order. The new kind of curvaton model was first proposed in Ref. \cite{Feng:2013pba}. The benefit of this model is that due to the coupling which contributes a factor of $H^2$ to the kinetic term, the perturbations of curvaton feel like in a nearly de-Sitter spacetime and will give rise to scale-invariant power spectrum favored by the data, independent of the details of the background evolution of the universe. Since the curvaton field couples nonminimally to gravity, despite of the pure scalar and tensor bispectra, the cross correlation of tensor (gravitational) and scalar (field) perturbations will give nontrivial contributions to the non-Gaussianities, so we perform a full calculation of all the 3-points correlation functions, and get all the possible shape functions.

However, the requirements of stabilities and gravitational waves do give certain constraints on the model. According to our previous study \cite{Feng:2013pba}, this model can work very well with the condition $|y|\ll M_p^2$, and can act as a low scale inflation in the expanding universe. According to our Eqs. (\ref{fnleql2}), the non-Gaussian estimators $f_{NL}$ is proportional to the ratio of $|y|/M_p^2$, and thus can give rise to small non-Gaussianities which is well within the strong constraint of PLANCK data. Our result shows that for modest parameter choices, $f_{NL}$ can be of ${\cal O}(0.1)$. This indicates that our model can be a viable model and can have very prosperous developments.

Other than $f_{NL}$'s of the pure scalar part, our model can also be tested by the observations on non-Gaussianities of pure tensor and mixed parts. In our model, the pure tensor and mixed parts could generate sizable $f_{NL}^{sqz}$ and $f_{NL}^{eql}$, respectively. If the future surveys can observe modest signals of squeezed non-Gaussianities of pure tensor perturbations, or equilateral non-Gaussianties of mixed tensor-scalar perturbations, it will be a good support of our model.

\section*{Acknowledgments}
We thank Yun-Song Piao for useful discussion. T.Q. also acknowledges Xian Gao and Gary Shiu for their helpful suggestions. The work of T.Q. is supported by the Open Project Program of State Key Laboratory of Theoretical Physics, Institute of Theoretical Physics, Chinese Academy of Sciences, China (No.Y4KF131CJ1). The work of K.F. is supported in part by NSFC under Grant No:11222546, in part by National Basic Research Program of China, No:2010CB832804.

\end{document}